\documentclass{emulateapj}
\usepackage[backref=section]{hyperref}

\usepackage[dvipsnames]{xcolor}
\colorlet{mylinkcolor}{Maroon}
\colorlet{mycitecolor}{MidnightBlue}
\colorlet{myurlcolor}{MidnightBlue}
\hypersetup{ 
	colorlinks   = true,
	citecolor    = mycitecolor,
	linkcolor    = mylinkcolor,
	urlcolor	= myurlcolor
}

\usepackage{etoolbox}
\makeatletter
\patchcmd{\BR@backref}{\newblock}{\newblock[}{}{}
\patchcmd{\BR@backref}{\par}{]\par}{}{}
\makeatother
\usepackage[all]{hypcap}

\usepackage{graphicx}
\usepackage{epsfig}
\usepackage{natbib}
\usepackage{multirow}
\usepackage{amsmath,amssymb}
\usepackage{rotating}
\usepackage{pdfpages}
\usepackage{xcolor}
\usepackage[flushleft]{threeparttable}
\usepackage{booktabs}

\usepackage{CJKutf8}
\usepackage{ucs}
\usepackage[encapsulated]{CJK}
\newcommand{\myfont}{bsmi} 

\def\gsim{\mathrel{\raise0.35ex\hbox{$\scriptstyle >$}\kern-0.6em
		\lower0.40ex\hbox{{$\scriptstyle \sim$}}}}
\def\lsim{\mathrel{\raise0.35ex\hbox{$\scriptstyle <$}\kern-0.6em
		\lower0.40ex\hbox{{$\scriptstyle \sim$}}}}

\submitted{{\it ApJ} in press}

\lefthead{Chen et al.}
\righthead{Counterpart identifications for SMGs in S2CLS-UDS}

\begin{document}

\title{The SCUBA-2 Cosmology Legacy Survey: Multi-wavelengths counterparts to 10$^3$ submillimeter galaxies in the UKIDSS-UDS field}

\begin{CJK}{UTF8}{\myfont} 
\author{Chian-Chou Chen  (陳建州)\altaffilmark{1,2}, Ian Smail\altaffilmark{1,2}, Rob J. Ivison\altaffilmark{3,4}, Vinodiran Arumugam\altaffilmark{3,4}, Omar Almaini\altaffilmark{5}, Christopher J. Conselice\altaffilmark{5}, James E. Geach\altaffilmark{6}, Will G. Hartley\altaffilmark{5,7}, Cheng-Jiun Ma\altaffilmark{1,2},  Alice Mortlock\altaffilmark{5,3}, Chris Simpson\altaffilmark{8}, James M. Simpson\altaffilmark{1,2}, A. Mark Swinbank\altaffilmark{1,2}, Itziar Aretxaga\altaffilmark{9}, Andrew Blain\altaffilmark{10}, Scott C. Chapman\altaffilmark{11}, James S. Dunlop\altaffilmark{3}, Duncan Farrah\altaffilmark{12}, Mark Halpern\altaffilmark{13}, Micha\l{} J. Micha\l{}owski\altaffilmark{3}, Paul van der Werf\altaffilmark{14}, Aaron Wilkinson\altaffilmark{5}, Jorge A. Zavala\altaffilmark{9}}

\altaffiltext{1}{Centre for Extragalactic Astronomy, Department of Physics, Durham University, South Road, Durham DH1 3LE, UK}
\altaffiltext{2}{Institute for Computational Cosmology, Durham University, South Road, Durham DH1 3LE, UK}
\altaffiltext{3}{Institute for Astronomy, University of Edinburgh, Royal Observatory, Blackford Hill, Edinburgh EH9 3HJ, UK}
\altaffiltext{4}{European Southern Observatory, Karl Schwarzschild Strasse 2, Garching, Germany}
\altaffiltext{5}{University of Nottingham, School of Physics and Astronomy, Nottingham, NG7 2RD UK}
\altaffiltext{6}{Center for Astrophysics Research, Science \& Technology Research Institute, University of Hertfordshire, Hatfield AL10 9AB, UK}
\altaffiltext{7}{ETH Z\"{u}rich, Institut f\"{u}r Astronomie, HIT J 11.3, Wolfgang-Pauli-Strasse 27, CH-8093 Z\"{u}rich, Switzerland}
\altaffiltext{8}{Astrophysics Research Institute, Liverpool John Moores University, Liverpool Science Park, 146 Brownlow Hill, Liverpool L3 5RF}
\altaffiltext{9}{Instituto Nacional de Astrof\'{i}sica, \'{O}ptica y Electr\'{o}nica (INAOE), Luis Enrique Erro 1, Sta. Ma. Tonantzintla, Puebla, Mexico}
\altaffiltext{10}{Physics \& Astronomy, University of Leicester, Leicester, LE1 7RH, UK}
\altaffiltext{11}{Department of Physics and Atmospheric Science, Dalhousie University, 6310 Coburg Rd., Halifax, NS B3H 4R2, Canada}
\altaffiltext{12}{Department of Physics, Virginia Tech, Blacksburg, VA 24061, USA}
\altaffiltext{13}{Department of Physics \& Astronomy, University of British Columbia, 6224 Agricultural Road, Vancouver, BC, Canada V6T 1Z1}
\altaffiltext{14}{Leiden Observatory, Leiden University, P.O. Box 9513, NL-2300 RA Leiden, The Netherlands}
\setcounter{footnote}{2}

\begin{abstract}
We present multiwavelength identifications for the counterparts of 1088 submillimeter sources detected at 850\,$\mu$m in the SCUBA-2 Cosmology Legacy Survey study of the UKIDSS-UDS field. By utilising an ALMA pilot study on a subset of our {bright} SCUBA-2 sample as a training set, along with the deep optical-near-infrared data available in this field, we develop a novel technique, Optical-IR Triple Color (OIRTC), using $z-K$, $K-[3.6]$, $[3.6]-[4.5]$ colors to select the candidate submillimeter galaxy (SMG) counterparts. By combining radio identification and the OIRTC technique, we find counterpart candidates for 80\% of the {\it Class} = 1 $\geq4\,\sigma$ SCUBA-2 sample, defined as those that are covered by both radio and OIR imaging and the base sample for our scientific analyses. Based on the ALMA training set, we expect the accuracy of these identifications to be $82\pm20$\%, with a completeness of $69\pm16$\%, {essentially as accurate as the traditional $p$-value technique but with higher completeness}. We find that the fraction of SCUBA-2 sources having candidate counterparts is lower for fainter 850\,$\mu$m sources, and we argue that for follow-up observations sensitive to SMGs with $S_{850}\gtrsim 1$\,mJy across the whole ALMA beam, the fraction with multiple counterparts is likely to be $>40$\% for SCUBA-2 sources at $S_{850} \gtrsim 4$\,mJy. We find that the photometric redshift distribution for the SMGs is well fit by a lognormal distribution, with a median redshift of $z=2.3\pm0.1$. After accounting for the sources without any {radio and/or OIRTC} counterpart, we estimate the median redshift to be $z=2.6\pm0.1$ for SMGs with $S_{850} >1$\,mJy. We also use this new large sample to study the clustering of SMGs and the the far-infrared properties of the unidentified submillimeter sources by stacking their {\it Herschel} SPIRE far-infrared emission.
\end{abstract}


\section{Introduction}

Ultra-luminous infrared galaxies (ULIRGs, with infrared luminosities of $L_{\rm IR}\geq $\,10$^{12}$\,L$_{\odot}$; \citealt{Sanders:1996p6419}) are relatively rare at $z\sim$\,0, but their space density rapidly increases with look-back time and apparently peaks around $z\sim$\,2--3 (e.g., \citealt{Barger:1999p6801, Chapman:2005p5778, Le-Floch:2005fk, Gruppioni:2013aa}). The vast majority of the luminosity of these sources escapes in the far-infrared (far-IR) and submillimeter and as a result they are the brightest extragalactic sources in the far-IR/submillimeter sky. The shape of the dust spectral energy distribution (SED) peaks around $\sim$\,100\,$\mu$m and declines at longer wavelengths (e.g., \citealt{U:2012aa, Symeonidis:2013aa, Swinbank:2014ul}). This characteristic form yields a ``negative $K$-correction'' for observations in the submillimeter waveband \citep{Blain:1993aa}, with the apparent flux of a source with a fixed infrared luminosity remaining almost constant over a wide range in redshift, $z\sim$\,1--6 (e.g., \citealt{Blain:2002p8120, Casey:2014aa}). When combined with the typical sensitivities and confusion limits of existing far-IR/submillimeter observatories (e.g., {\it Herschel}, the James Clerk Maxwell Telescope [JCMT] or the Large Millimetre Telescope [LMT]), the negative $K$-correction means that surveys for high-redshift ULIRGs are most efficiently undertaken in wavebands around $\sim$\,1\,mm, leading to the association of name ``submillimeter galaxies'' (SMGs) with this population (e.g., \citealt{Smail:1997p6820}). Moreover, the surface density of high-redshift ULIRGs is also best-matched to the wide-field capabilities of single-dish telescopes, rather than the narrow field of view of current (sub)millimeter interferometers.  This has been the motivation for a series of panoramic (sub)millimeter surveys over the past decade using first-generation bolometer cameras on the JCMT, the IRAM 30-m, APEX and ASTE (e.g., \citealt{Barger:1998p13566, Hughes:1998p9666, Scott:2002p6539, Coppin:2006p9123, Bertoldi:2007aa, Weis:2009qy, Scott:2010pp, Ikarashi:2011aa}). By exploiting the technical advances in the fabrication of bolometer cameras, specifically the SCUBA-2 camera on JCMT, recent submillimeter surveys have been mapping the sky in an unprecedented speed \citep{Chen:2013fk, Chen:2013gq, Geach:2013kx, Casey:2013aa}. Most recently an international team completed the SCUBA-2 Cosmology Legacy Survey (S2CLS), undertaking panoramic surveys on square degree areas down to mJy sensitivity limits (see Geach et al.\ in prep for a description of the survey).   

To use these (sub)millimeter surveys to understand the cause of the rapid evolution of the ULIRG population and its relation to the galaxy populations seen both today and at earlier times, it is essential to reliably locate the counterparts to the (sub)millimeter sources at other wavelengths, necessary to understand the physical properties and astrophysics of these systems. However, the combination of the high dust obscuration in these systems, their high redshifts, and the coarse resolution of the (sub)millimeter maps ($\gsim $\,10--30$''$ FWHM) provided by single-dish observatories make this process challenging. Much of the early work on the properties of high-redshift ULIRGs relied on identifications based on indirect tracers of the far-IR/submillimeter emission such as the radio, near-IR, and mid-IR (e.g., \citealt{Ivison:1998p10286, Ivison:2002uq, Smail:2000p6377};  although see \citealt{Downes:1999aa, Dannerbauer:2002aa}). These techniques have been used to derive identifications for samples of $\sim$\,100 SMGs from a number of surveys (e.g., \citealt{Pope:2006p8076, Lindner:2011aa, Biggs:2011uq, Yun:2012aa, Michaowski:2012aa, Alberts:2013aa}), but are known to be biased against identifying the highest-redshift ULIRGs due to the absence of a negative $K$-correction in the radio or IR (e.g., \citealt{Chapman:2005p5778}). 

Luckily, in parallel with the developments of new large-format (sub)millimeter bolometer cameras, the commissioning of the Atacama Large Millimeter Array (ALMA) and upgrades to the Submillimeter Array (SMA) and the Northern Extended Millimetre Array (NOEMA) have produced a similar advance in the capabilities of (sub)millimeter interferometers for studying submillimeter sources (e.g., \citealt{Gear:2000aa, Dannerbauer:2002aa, Dannerbauer:2008aa, Iono:2006p6985, Tacconi:2006p8449, Wang:2007p6971, Wang:2011p9293, Younger:2007p6982, Younger:2008p8372, Younger:2009p9502, Cowie:2009p6978, Aravena:2010p8370, Tamura:2010aa, Knudsen:2010lr, Chen:2011p11605, Chen:2014aa, Barger:2012lr, Barger:2014aa, Smolcic:2012pp, Hodge:2013lr, Simpson:2015aa, Simpson:2015ab, Ikarashi:2014aa, Miettinen:2015aa}).  

The first results from ALMA on the identifications of SMG counterparts to submillimeter sources by \citet{Hodge:2013lr} confirmed some of the biases and incompleteness arising from the use of radio and mid-IR, with $\sim$80\% of SMGs correctly identified but with a completeness of just 45\%. While such interferometric studies highlight the usefulness of obtaining identifications in the submillimeter for SMGs, the limited time available on these facilities means it is currently {time-expensive} to use them to map the very large samples of (sub)millimeter sources from the latest bolometer surveys. For this reason we are driven back to using the indirect tracers. However, techniques that are used to select the counterpart candidates can be investigated, trained, and improved by using the results of these interferometric observations.

Here we present counterparts to the $\sim$\,1,000 submillimeter sources that have been detected in the S2CLS 850\,$\mu$m map of the UKIDSS Ultra-Deep Survey field (UDS). This is one of the widest and most sensitive blank-field 850\,$\mu$m surveys yet completed, with a median 1\,$\sigma$ noise of 1\,$\sigma = 0.9$\,mJy across $\sim$1 degree$^2$ (Geach et al.\ in preparation). The UKIDSS-UDS field is an exceptionally well-studied region of the extragalactic sky with sensitive multi-wavelength coverage of the $\sim$1 degree$^2$ region from the ultraviolet to mid-infrared and radio (see \autoref{sec:obs} for references). Our analysis also takes advantage of deep ALMA Cycle 1 observations of a subset of the submillimeter sources in this region \citep{Simpson:2015ab, Simpson:2015aa}, which provide an invaluable resource for training and developing new techniques to select the SMG counterparts to the submillimeter sources detected in the low-resolution single-dish observations.

The structure of the paper is: in \S2 we describe the submillimeter survey of the UKIDSS-UDS field, along with the ancillary data from X-ray to radio that are used in this study.  \S3 then describes the process, including the novel OIRTC technique, developed for the identifications, which exploits the ALMA identifications of SMGs associated with a sample of the brighter submillimeter sources in the field. In \S4 we present the catalog of counterpart candidates and discuss their properties, while \S5 gives our summary. 
Throughout this paper we adopt the AB magnitude system \citep{Oke:1983aa}, and we assume the {\it Planck} cosmology: H$_0 =$\,67.77\,km\,s$^{-1}$ Mpc$^{-1}$, $\Omega_M = $\,0.31, and $\Omega_\Lambda =$\,0.69 \citep{Planck-Collaboration:2014aa}.

\begin{figure*}
\begin{center}
    \leavevmode
      \includegraphics[scale=1]{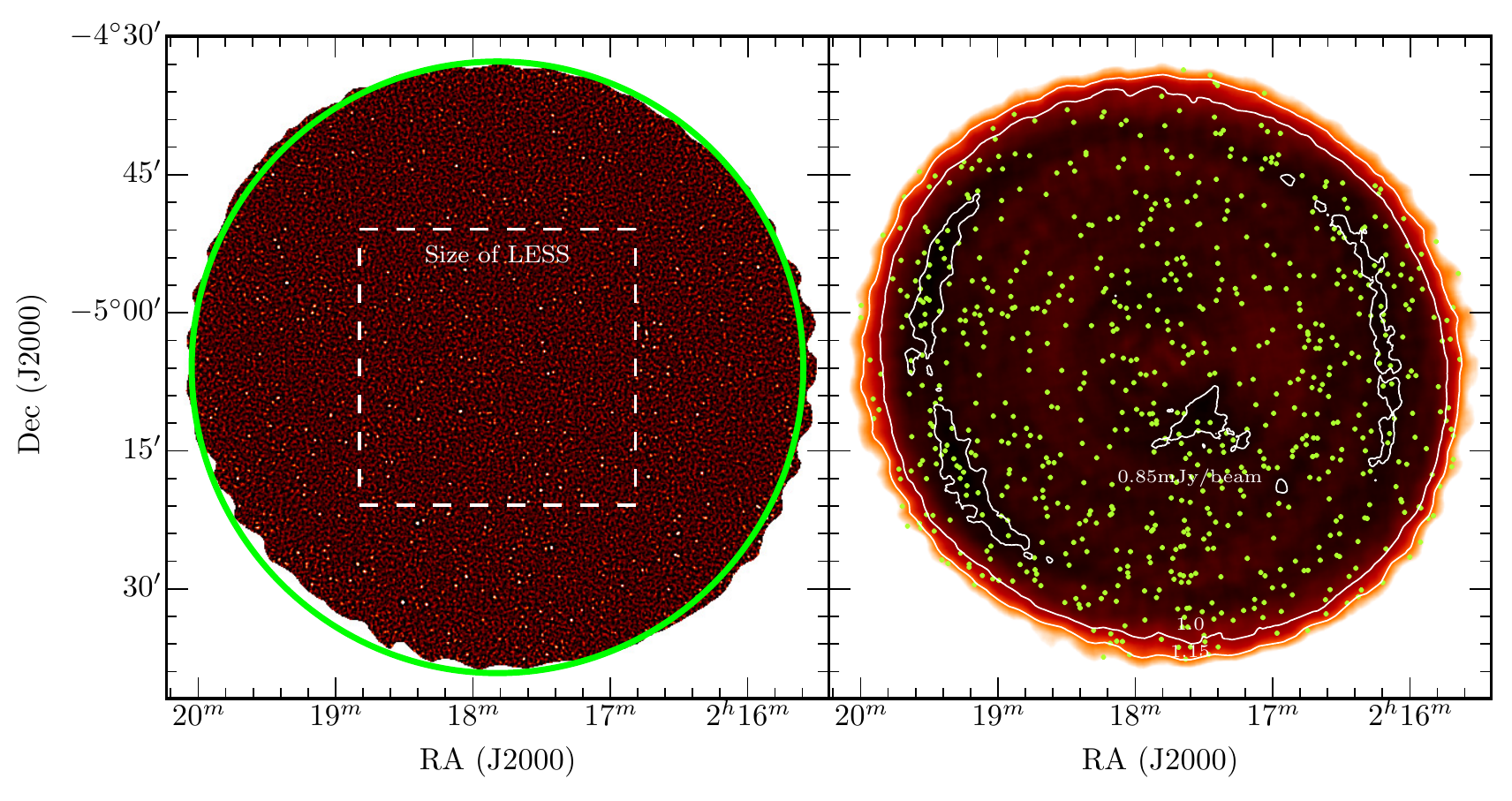}
       \caption{An overview of S2CLS-UDS. {\it Left:} The matched-filtered S2CLS SCUBA-2 850\,$\mu$m flux density map of the UKIDSS UDS field, linearly scaled between -1 to 5\,mJy. The green circle roughly outlines the survey area with a $\sim$1.1 degree diameter, $\sim4\times$ larger than the previously largest 850\,$\,\mu$m uniform survey in a single field -- the LABOCA survey in the ECDF-S (LESS; \citealt{Weis:2009qy}). For comparison, the size of LESS is shown in the white dashed box. {\it Right:} A color plot showing the r.m.s. values of the matched-filtered S2CLS map, linearly scaled between 0.82 and 1.3\,mJy beam$^{-1}$. White contours are at 0.85, 1.0, and 1.15\,mJy beam$^{-1}$. Green points mark our 716 $\geq$\,4\,$\sigma$ detections in the {\sc main} sample. Our SCUBA-2 map is $\sim$40\% deeper in sensitivity and has spatial resolution $\sim$30\% higher compared to LESS, yielding a SMG sample $\sim6\times$ bigger than the LESS survey and making the S2CLS-UDS the largest uniform 850\,$\mu$m survey by far.
       }
    \label{s2clsuds}
 \end{center}
\end{figure*}

\section{Observations, reduction and supporting data}\label{sec:obs}

\subsection{SCUBA-2}

The SCUBA-2 data at 850\,$\mu$m in the UDS field were taken as part of the S2CLS. The full data reduction steps are described fully in Geach et al.\ (2015 in preparation), but we describe the main steps here.  The Dynamical Iterative Map-Maker ({\sc dimm}) within the {\it Sub-Millimetre Common User Reduction Facility} ({\sc smurf}; \citealt{Chapin:2013fk}) is used to extract astronomical signal from each SCUBA-2 bolometer time stream, mapping the result onto a celestial projection. All S2CLS maps are projected on a tangential co-ordinate system with 2$''$ pixels.

Flat-fields are applied to the time-streams using flat scans that bracket each observation, and a polynomial baseline fit is subtracted from each time stream. Data spikes are rejected (using a 5\,$\sigma$ threshold in a box size of 50 samples), DC steps are removed and gaps filled. Next, an iterative process begins that aims to fit the data with a model comprising a common mode signal, astronomical signal and noise. The common mode modelling is performed independently for each SCUBA-2 sub-array, deriving a template for the average signal seen by all the bolometers; it is removed from the stream, and an extinction correction is applied \citep{Dempsey:2013qy}. Next, a filtering step is performed in the Fourier domain, which rejects data at frequencies
corresponding to angular scales $\theta>150''$ and $\theta< 4''$. Finally, a model of the  astronomical signal is determined by gridding the time streams onto a celestial projection (since a given sky position will have been visited by many independent bolometers) and then subtracted from the input time streams. The iterative process continues until the residual between the model and the data converges.

The last processing step is to apply a matched filter to the maps, convolving with the instrumental PSF to optimize the detection of point sources. We use the {\sc picard} recipe {\it scuba2\_matched\_filter} which first smooths the map (and the PSF) with a 30$''$ gaussian kernel, then subtracts this from both to remove any large scale structure not eliminated in the filtering steps that occurred during the {\sc dimm} reduction. The map is then convolved with the smoothed beam. A flux conversion factor of 591\,Jy\,beam$^{-1}$\,pW$^{-1}$ is applied; this canonical calibration is the average value derived from observations of hundreds of standard submillimetre calibrators observed during the S2CLS campaign \citep{Dempsey:2013qy}, and includes a 10\% correction necessary to account for losses that occur due to the combination of filtering steps we apply to the data (see \citealt{Geach:2013kx}). The flux calibration is expected to be accurate to within 15\%.

The final matched-filtered map has a noise of 0.82\,mJy beam$^{-1}$ at the deepest part and better than $\le 1.3$\,mJy rms over a $\sim$\,1.0 degree$^2$ (a $\sim$1.1 degree diameter circle). The coverage is relatively uniform and the median depth within this region is 0.89\,mJy beam$^{-1}$  (\autoref{s2clsuds}). 

We apply a simple source detection and extraction algorithm, described in more detail in Geach et al. (2015 in prep). In brief, we apply a top-down detection algorithm, first identifying the peak-pixel in the SNR map, recording position and flux and instrumental RMS and then subtracting a peak-scaled model of the PSF at this position. The next peak value is identified and the process repeated until a floor SNR threshold is met, which we set to 3.5\,$\sigma$.

In total we detect 1088 submillimeter sources at $\geq$ 3.5\,$\sigma$ within the region where rms noise is $\le $\,1.3\,mJy beam$^{-1}$. We define a {\sc main} sample of 716 submillimeter sources that have $\ge$ 4.0\,$\sigma$, for which we expect a false detection rate of $\sim$\,1\% based on simulations and source extractions on negative signals (Geach et al.\ in prep.). We also define a {\sc supplementary} sample of 372 submiliimeter sources that are detected at $3.5-4.0$\,$\sigma$ and have a false detection rate of $\sim$10\%. In this paper, we provide counterpart candidates for both {\sc main} and {\sc supplementary} samples, however the scientific analyses were performed on the {\sc main} sample. 

The maps are corrected for astrometry by adopting a shift of 0$\farcs$67 in R.A.\ and $-$2$\farcs$33 in Decl., based on stacking of the 850\,$\mu$m maps at the location of the radio sources. We have also stacked on the 850\,$\mu$m maps centered on the MIPS and $K$-band sources and found consistent results.

\subsection{ALMA}
We have carried out ALMA follow-up observations at 870\,$\mu$m on 30 of the brighter SCUBA-2 sources in a Cycle-1 project 2012.1.00090.S \citep{Simpson:2015ab, Simpson:2015aa}. These sources were selected to have $S_{850} \geq 8$\,mJy from an earlier version of the S2CLS map, but the sensitivity of the map has since improved and as a result 27 of the 30 ALMA targets still remain in our {\sc main} sample. Two of the three ALMA-observed sources that fall out of our {\sc main} sample (UDS252, UDS421) have no detection in the high-resolution ALMA observations but both are detected in {\it Herschel}/SPIRE imaging, while the remaining one (UDS298) is just below the 3.5\,$\sigma$ cut and the SCUBA-2 flux is consistent with the integrated flux of the two ALMA detections \citep{Simpson:2015ab}. With a median rms of $\sigma = $\,0.26\,mJy\,beam$^{-1}$, the primary ALMA catalog consists of 52 SMGs detected by ALMA at $> 4\,\sigma$, with a synthesized beam of $\sim $\,0$\farcs$8 FWHM. Higher-resolution versions of the maps, with $\sim$\,0$\farcs$3 FWHM, were used by \citet{Simpson:2015aa} to study the  sizes and light profile of the brighter SMGs at 870\,$\mu$m, while the descriptions of the bright source counts and the data reduction and source extraction can be found in \citet{Simpson:2015ab}. In this paper, we use these ALMA-detected SMGs as the training set to formulate our methodology to identify candidates counterpart for the rest of the SCUBA-2 SMG sample. Note that although the ALMA observations were conducted at a slightly different wavelength compared to the selection wavelength from SCUBA-2 (870\,$\mu$m versus 850\,$\mu$m), the difference in flux measurements is expected at $\sim$5\% level, which is negligible compared to the flux calibration error. Throughout this paper, we therefore denote $S_{850}$ as the fluxes that are measured at both 850\,$\mu$m and 870\,$\mu$m.

\subsection{Multi-wavelength ancillary data}\label{sec:ancillary}

\begin{figure}
\begin{center}
    \leavevmode
      \includegraphics[scale=0.85]{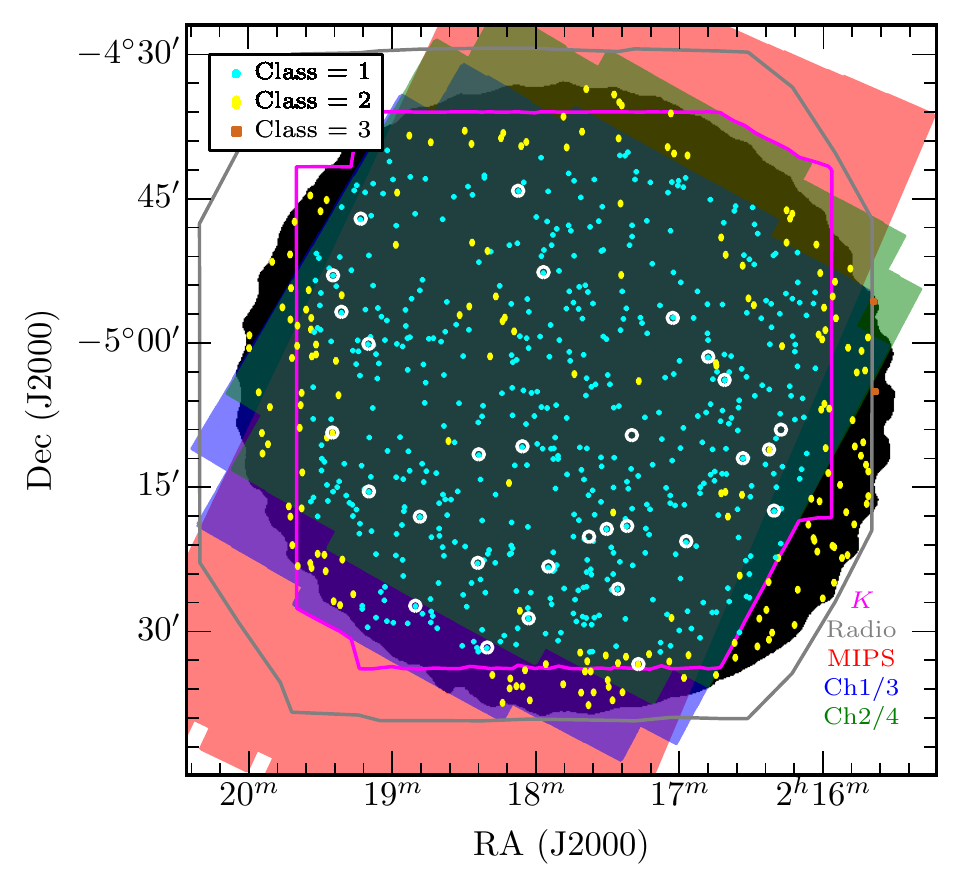}
       \caption{Multi-wavelengths coverage of $K$-band, {\it Spitzer}, and VLA in the UDS field, overlaid on the SCUBA-2 field shown as the black background. Similar to \autoref{s2clsuds}, the points mark the positions of our 716 $\geq4.0\,\sigma$ {\sc main} SCUBA-2 sample, while those enclosed with white circles were observed in our Cycle 1 ALMA program. Among the {\sc main} sample, all but the right-most two in the figure (brown; $Class =3$ sources) are covered by the radio imaging, and majority (73\%) of them have optical-near-infrared coverage suitable for our novel OIRTC technique (cyan; {\it Class} = 1 sources), with at least two color measurements available among $z-K$, $K-[3.6]$, and $[3.6]-[4.5]$ (see \autoref{subsec:oir}). The yellow points are {\it Class} = 2 sources that are covered by the radio imaging but not suitable for the OIRTC technique (more about classifications see \autoref{subsec:id}). 
}
    \label{multicover}
 \end{center}
\end{figure}

The $\sim$\,1 square degree UDS field contains a rich set of ancillary data. \autoref{multicover} roughly outlines the coverage of each indicated waveband. 

The $K$-band based multi-wavelength photometry adopted in this paper is based on the UDS data release 8 (DR8) of the UKIRT Infrared Deep Sky Survey (UKIDSS; \citealt{Lawrence:2007aa}). The UDS field is the deepest of the five sub-surveys of UKIDSS, consisting of four Wide-Field Camera (WFCAM; \citealt{Casali:2007aa}) pointings, covering 0.77 square degrees in $J$, $H$ and $K$ bands. The DR8 release contains all UDS data taken from 2005 to 2010. The 5\,$\sigma$ median depths are $J = $\,24.9, $H =$\,24.2 and $K =$\,24.6 (in a 2$''$ diameter aperture). Detailed descriptions of mosaicing, catalogue extraction and depth estimation will be presented in Almaini et al.\ (in preparation). After masking bad regions, removing bright stars and image artefacts produced by amplifier cross-talk, a $K$-band parent sample of a total of 159,871 sources was constructed for our analyses. 

The UDS field was observed by the Subaru telescope using the Suprime-Cam in five broadband filters, $B$, $V$, $R_c$, $i'$, and $z'$, to the limiting depths of $B =$\,28.4, $V =$\,27.8, $R_c =$\,27.7, $i' =$\,27.7, and $z' =$\,26.6, respectively (3\,$\sigma$, 2$''$ diameter apertures). Details of the Suprime-Cam survey are provided in \citet{Furusawa:2008aa}.
The field was also covered by the Megacam $u'$-band on the Canada-France-Hawaii Telescope (CFHT), with a 5\,$\sigma$ depth reaching $u'=$\,26.75 in a 2$''$ diameter aperture. The X-ray data were obtained as part of the Subaru-{\it XMM/Newton} Deep Survey (SXDS), consisting of seven contiguous fields with a total exposure of 400\,ks in the 0.2--10 keV band \citep{Ueda:2008aa}. Finally, the UDS field was imaged in mid-infrared with IRAC and 24\,$\mu$m MIPS by the {\it Spitzer} Legacy Program SpUDS. SpUDS data reach 5\,$\sigma$ depths of 24.2 and 24.0 AB magnitude at 3.6 and 4.5\,$\mu$m. 

Eleven-band photometry ($UBVRIzJHK$[3.6][4.5]) was measured with 3$''$ diameter apertures placed on each aligned image at the position of the $K$-band sources, {motivated by the fact that $K$-band is generally a good stellar mass indicator that is less affected by dust compare to other optical/NIR bands, with a data quality that is deeper and has a higher angular resolution compare to that of the IRAC bands}. To account for the correlated noise that is not represented in the weight maps, the magnitude uncertainties estimated by {\sc sextractor} are corrected by scaling the weight maps such that the uncertainty in source-free regions matches the rms measured from apertures placed on the science image. Three of the bands (the CFHT $u'$ band and the two IRAC channels) required aperture corrections to their photometry in order to obtain correct colors. This correction was performed based on smoothing the $K$-band images to the appropriate PSF and re-computing the aperture photometry to evaluate the expected changes. More details can be found in \citet{Hartley:2013aa}.

Photometric redshifts ($z_{\rm photo}$) have been derived for the DR8 parent sample, and the full description can be found in \citet{Hartley:2013aa} and \citet{Mortlock:2013aa}. In summary, the photometric 
redshifts are estimated using the {\sc eazy} template-fitting package \citep{Brammer:2008aa} through a maximum likelihood analysis. The default set of six templates does not sufficiently represent all of our galaxies, in particular the $u'$-band flux is significantly overestimated on the blue objects at high redshift. A seventh template is therefore constructed by applying a small amount of Small Magellanic Cloud-like extinction \citep{Prevot:1984aa} to the bluest template in {\sc eazy}. 

To assess the accuracy of these photometric redshifts and to determine the cut on the $\chi^2$ from the template-fitting, we compare the derived values to the spectroscopic redshifts ($z_{spec}$) that are available in the UDS. A large fraction of these $z_{spec}$ came from the UDSz, a European Southern Observatory large spectroscopic survey (ID:180.A-0776; Almaini et al., in preparation) and also from the literature (see \citealt{Simpson:2012ab} and references therein). After excluding bright X-ray and radio sources that are likely to be AGNs \citep{Simpson:2006aa, Ueda:2008aa}, there are 2,745 sources with measured spectroscopic redshifts ($z_{\rm spec}$). If we only consider 2,461 sources that have $\chi^2 < 10$ in the $z_{\rm photo}$ fitting, we find a dispersion in $(z_{\rm photo}-z_{\rm spec})/(1+z_{\rm spec})$, after excluding outliers ($\Delta z/(1+z_{\rm spec}) > 0.15$; $<$\,4\%), is $\Delta z/(1+z_{\rm spec}) \sim$\,0.02, slightly better than what was found in \citet{Hartley:2013aa}. We therefore apply a $\chi^2$ cut of ten in this paper.

For the 24-$\mu$m MIPS image, we use {\sc SExtractor} \citep{Bertin:1996zr} to extract sources. Following \citet{Shupe:2008aa}, the local background is estimated using a box with 128\,$\times$\,128 pixels with a pixel size of 1$\farcs$2, which then is used to weight the source extraction. Given MIPS has a beam FWHM of $\sim$\,6$''$ at 24\,$\mu$m, we expect most extragalactic sources to be unresolved and so appear as point sources. We thus set the detection threshold to be 20 connected $\geq $\,2\,$\sigma$ pixels (1 beam area). This selection corresponds to a $\sim$\,4\,$\sigma$ point source detection. In total we detect 12,127 24\,$\mu$m sources within the SCUBA-2 coverage. We estimate the number of false detections by inverting the map and extracting the negative sources using the same detection parameters. We find the false detection rate to be 0.03$^{+0.04}_{-0.02}$\%, consistent with a $\sim$\,4\,$\sigma$ detection, which increases to 0.1\% if we account for sources close to the edge of the 24\,$\mu$m map. We measure photometry in a 15$''$-diameter 
aperture. We estimate the aperture correction by median stacking the unsaturated bright stars and compute the curve of growth. We find an aperture correction factor of 1.5, consistent with the MIPS maps in the SWIRE fields \citep{Shupe:2008aa}. The 1\,$\sigma$ uncertainty is 24\,$\mu$Jy, estimated using a 7$\farcs$5-radius 
aperture in the source-free regions.

The VLA radio observations at 1.4\,GHz (20\,cm) were carried out by the project UDS20 (Arumugam et al. in preparation), which comprises a mosaic of 14 pointings covering a total area of $\sim$\,1.3 degrees$^2$ centered on the UDS. All but two $\geq4\,\sigma$ SCUBA-2 sources are covered by the VLA map (\autoref{multicover}). The total integration time is $\sim$\,160\,hrs in A, B, and C--D configurations, yielding a nearly constant rms noise of $\sim$\,10\,$\mu$Jy across the full field ($< $\,8$\mu$Jy at the field center) and a beam size of 1$\farcs$8 FWHM. A total of $\sim$\,7,000 sources are detected above 4\,$\sigma$.
The full descriptions of the observations, data reductions, and the catalog are presented in V.\ Arumugam et al.\ (in preparation). 

%
%

\section{Counterpart identification}

%
%
\begin{figure*}
\begin{center}
    \leavevmode
  \includegraphics[scale=1]{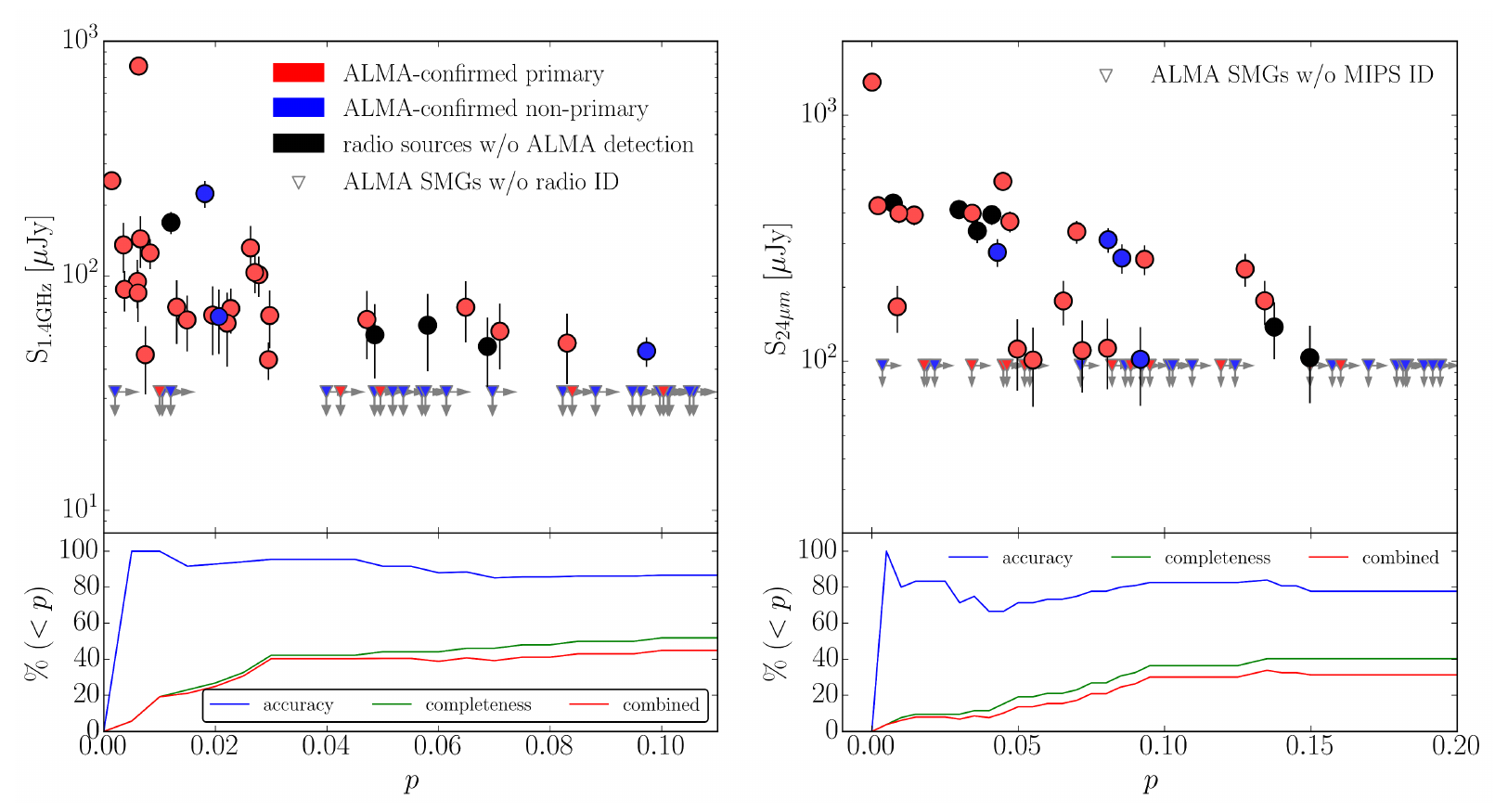}
       \caption{{\it Left:} The upper panel shows a plot on the radio flux versus the $p$-value of all 52 ALMA-detected SMGs in the UDS field \citep{Simpson:2015ab}, with upper limits given for SMGs without radio counterparts to within 1$\farcs$5. We identify where the SMGs  are the primary source, meaning they are the brightest sources if there are multiple detections in the ALMA maps, as well as where they are fainter than the brightest SMG in the map. Single detections are counted as primaries, and we also indicate the radio sources that are not detected in the ALMA imaging.
       The lower panel shows the accuracy and completeness of the identifications to the whole 52 SMGs 
       and the combined product of accuracy and completeness for sources with $p$ lower than specified values on the abscissa. {\it Right:} Same as the left panels, but on MIPS 24\,$\mu$m sources}
    \label{pstat}
 \end{center}
\end{figure*}

In this section, we utilise the sample of 52 ALMA-detected SMGs with S$_{850} \gsim 1.0$\,mJy from \citet{Simpson:2015aa, Simpson:2015ab} found in the vicinity of 30 of the brighter SCUBA-2 sources in the UDS to test various counterpart identification methods that are widely used in the literature. We also use this training set to develop a novel optical-near-infrared color method to supplement the traditional radio selection. We then apply the counterpart identification methodology to the whole sample of SCUBA-2 submillimeter sources. 

The main parameters we consider in the tests and the training are accuracy and completeness, which are defined as

\begin{equation*}
\begin{split}
Accuracy = \frac{N_{\rm confirmed}}{N_{\rm selected}} \times 100\%\\
Completeness = \frac{N_{\rm confirmed}}{N_{\rm total}} \times 100\%, 
\end{split}
\end{equation*}
where $N_{\rm selected}$ is the number of selected candidate counterparts based on the selection methods, $N_{\rm confirmed}$ is the number of selected candidates that are actually confirmed by ALMA based on the training set, and $N_{\rm total} = 52$ representing the total number of the ALMA-detected SMGs. The decision of the best strategy is made by maximising the product of both parameters, and the quoted errors are Poisson if not specifically stated.

\subsection{Radio identifications}\label{subsec:radio}

%
%
\begin{figure*}
\begin{center}
    \leavevmode
      \includegraphics[scale=1]{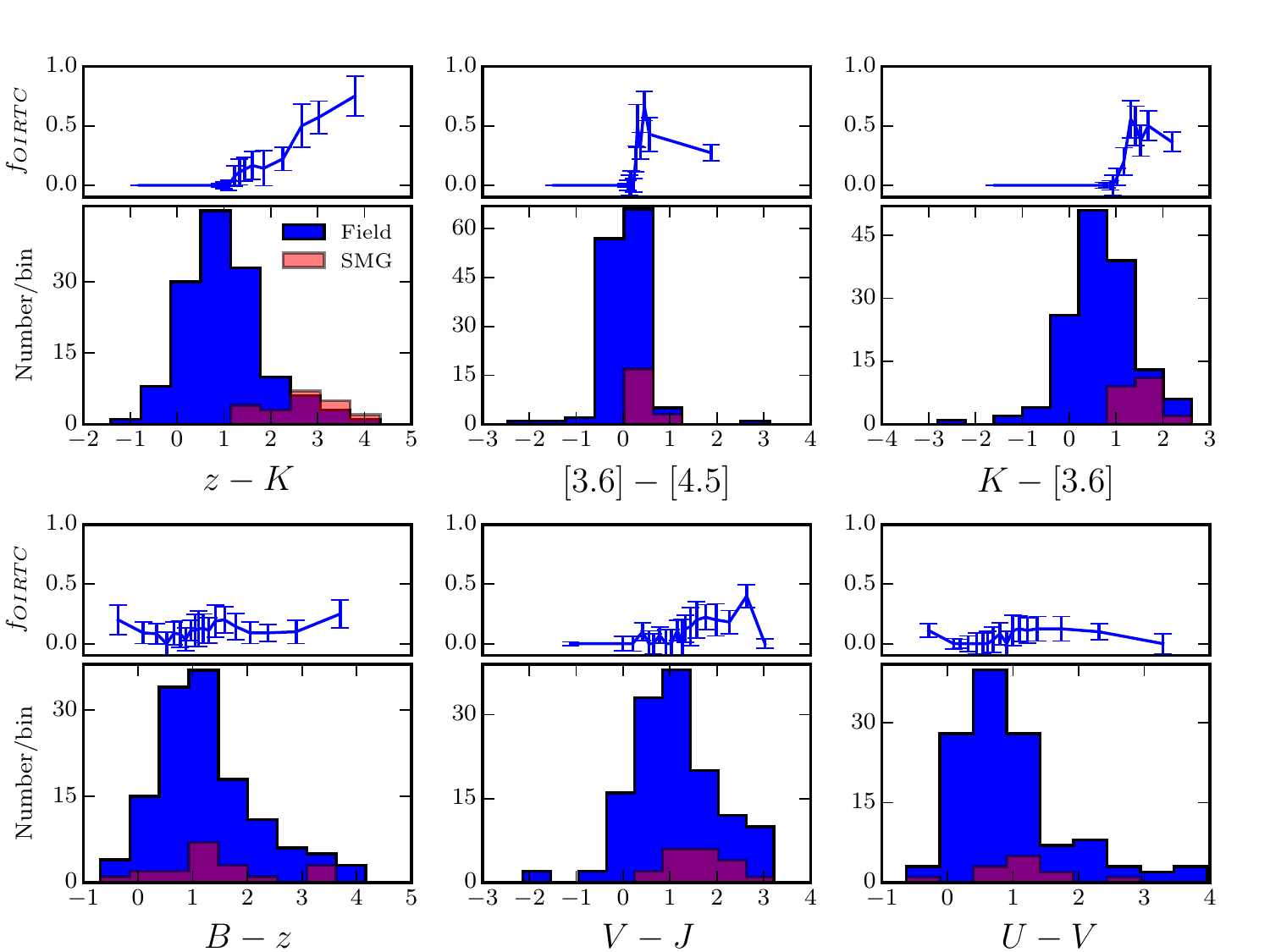}
       \caption{{\it Lower section in each sub-panel}: Histrograms in each specified color of the training sample, which are 164 $K$-band sources that are located within the primary beam of our 30 ALMA observations in the UDS \citep{Simpson:2015ab}, with red representing the 22 SMGs with $S_{850} > $\,2.7\,mJy that have matches to the $K$-band sources within 1$''$. The non-SMG field sources are shown in blue. {\it Upper panels in each sub-panel}: The SMG fraction as calculated by dividing number of SMGs to the total number of sources in each color bin. The errors are estimated through Monte Carlo simulations, in which we derive standard deviation of SMG fraction with 100 realisations of randomly populated data points based on their measured colors and errors. Distinct color distributions between SMGs and field sources are found in $(z-K)$, $(K-[3.6])$, and $([3.6]-[4.5])$, which are used to develop our OIRTC technique (\autoref{subsec:oir}). At a typical SMG redshift, $z\sim$\,2, these colors correspond to roughly rest frame $(U-R)$, $(R-J)$, and $(J-H)$, suggesting both the Balmer/4000\AA\, break and dust extinction could be the cause of SMGs being red in these colors.  
       }
    \label{colorhist}
 \end{center}
\end{figure*}

We first test the use of radio sources to locate SMGs associated with submillimeter sources selected from low-resolution, single-dish submillimeter surveys  (e.g., \citealt{Ivison:1998p10286, Ivison:2002uq, Ivison:2005aa, Lindner:2011aa}). This approach utilises the corrected-Poissonian probability, or the $p$-values, to estimate the likelihood of radio sources being a random chance association to the submillimeter sources. The calculation of the $p$-value is described in \citet{Downes:1986aa} as

\begin{equation}
p=1-exp(-\pi n \theta^2)
\end{equation}
where $n$ represents the radio source density and $\theta$ is the angular offset between the radio and the submillimeter source. A matched is typically considered reliable if $p < 0.05$ (e.g., \citealt{Ivison:2002uq, Pope:2006p8076, Chapin:2009aa, Yun:2012aa})

We investigate the accuracy and completeness of the radio counterpart identifications for ALMA-detected SMGs that are located within the ALMA primary beam (8$\farcs$7). To account for all possible counterparts to the single-dish submillimeter sources, we do not scale our search radius as a function of the SCUBA-2 detection signal-to-noise ratio (SNR) in calculating $p$-values, as has been done in some previous work (e.g., \citealt{Biggs:2011uq}). This is motivated by studies showing that due to the fact that single-dish detected SMGs tend to break into multiple sub-components in high-resolution follow-up observations, the separation between the sub-components and the corresponding single-dish source does not correlate with the SNR of the single-dish detection \citep{Hodge:2013lr}. This result suggests that employing a fixed search radius, instead of a SNR dependent radius, during the process of identifying candidate counterparts may be a better strategy. Note though, \citet{Simpson:2015ab} found that by convolving the ALMA maps with the SCUBA-2 beam, the radial separation between the convolved ALMA map centroid and the SCUBA-2 source is indeed a function of the SNR of the SCUBA-2 detection and consistent with Gaussian distribution.

Among the 52 ALMA SMGs in the training set, we found 27 that have radio counterparts matched to 1$\farcs$5. While all 27 of them have $p < $\,0.1, 23 have $p < $\,0.05 (the canonical value used in the literature to select ``robust'' SMG counterparts). On the other hand, if we look at all the 30 radio sources located within the ALMA primary beam, 24 of them have $p < $\,0.05. As a result, the accuracy of identifying SMG counterparts using radio sources with $p < $\,0.05 is 92$^{+8}_{-27}$\%(22/24)\footnote[15]{Two ALMA SMGs, UDS\,156.0 and UDS\,156.1, are matched to the same radio source.}, and that by using all radio sources (in this case $p < $\,0.1) is only slightly lower at 87$^{+13}_{-23}$\% (26/30). In fact, the accuracy ranges between 85\% to 100\% if we adopt any choice of $p$ below $p=$\,0.1 as the selection criterion, with no statistical difference (\autoref{pstat}). Although the accuracy is indeed lower for radio identified counterpart with $p=$\,0.05--0.1 at 66$^{+34}_{-43}$\%, the result  suffers from small number statistics and the difference is insignificant. In addition, these high-$p$ sources are generally located further away from the pointing center (the centroid of the SCUBA-2 source), and the decrease in ALMA sensitivity for these sources due to primary beam coverage could be the cause of this slight but insignificant drop in accuracy.

On the other hand, we find that the majority (88$^{+12}_{-25}$\%; 23/26) of the ALMA-confirmed radio identifications are the primary SMGs in the ALMA maps, defined as those brightest ALMA detections, which are found to dominate and contribute on average $\sim$\,75\% of the total flux measured by SCUBA-2 \citep{Simpson:2015ab}, although 18$\pm$9\% (5/28) of the primaries are not detected in the radio imaging.

In summary, 
we conclude that, at $p<0.1$ the accuracy of the radio identifications does not appear to depend on the $p$-value, and taking all radio sources within the ALMA primary beam as the SMG counterparts actually yields better completeness (27 out of 52; 52$\pm$12\%) and an overall identification performance by maximising the product of accuracy and completeness.  

\subsection{MIPS 24$\mu$m identifications}\label{subsec:mips}

We conduct a similar test of the $p$-value method using the 24-$\mu$m MIPS counterparts. We found that of the 52 ALMA SMGs there are 21 that have MIPS counterparts matched to within 2$''$, and among them 10 have $p < $\,0.05. There are 27 MIPS sources in total located within the ALMA primary beams of the 30 submillimeter sources, 14 of which have $p < $\,0.05. The percentage of $p < $\,0.05 MIPS sources which are confirmed SMGs is 71$^{+29}_{-30}$\% (10/14), and that of all MIPS sources is 78$^{+22}_{-23}$\% (21/27). This is a slightly lower rate than for the radio, reflecting the different strengths of the correlations between the radio and mid-IR emission to the far-IR/submillimeter, as well as the differing levels of contaminations from foreground populations.  Moreover, the FWHM of the 24\,$\mu$m MIPS images is 6$''$, much worse than that of the radio maps, and in this case source blending becomes an issue. Deciding to what extent to match MIPS sources to the ALMA SMGs is not straightforward. By expanding the matching radius to 3$''$, almost all (24/27) MIPS sources are matched to at least one ALMA-identified SMG. However, we find that by detailed comparison of the images some of these MIPS sources are not correct counterparts. For this reason we chose 2$''$ as a good balance to match most of the obviously correct counterparts without including many spurious ones. In the right panels of \autoref{pstat} we show that, similar to the radio counterparts, the accuracy of the MIPS counterparts does not depend on the $p$ values, and again, the completeness is significantly improved if one includes all MIPS sources that are located within the ALMA primary beam. We stress that changing the matching radius does not affect this result. 

\subsection{Optical-IR Triple Color (OIRTC)}\label{subsec:oir}

Previous studies have shown that SMGs are in general red in optical-near-infrared (OIR) colors such as $i-K$, $J-K$(DRGs), $K-[4.5]$ (KIEROs) (e.g., \citealt{Smail:2002p6793, Dannerbauer:2004aa, Frayer:2004fk, Wang:2012ys}), suggesting high-$z$ and dusty nature. OIR color cuts had therefore been used to identify potential counterparts (e.g., \citealt{Michaowski:2012aa}). However, while adopting single color cuts might select SMGs, the contaminations from the field sources can also be large. Color-color cuts, or characteristic density distribution, based on the {\it Spitzer} mid-IR observations have been proposed and used \citep{Yun:2008vn, Alberts:2013aa, Umehata:2014aa}. However, the training set for the mid-IR color-color techniques are heterogeneous, usually mixing with radio-, CO-, or SMA-identified SMGs, and the true accuracy and completeness of each technique are hard to understand. Armed with our ALMA data, which are based on a flux-limited SCUBA-2 sample, we can start looking into the best method using OIR colors to select SMGs counterparts. 

To separate SMGs from non-SMG field galaxies using multi-wavelength photometry, we constructed a training set based on the results of our ALMA follow-up observations, which targeted 30 brighter SMGs selected in the UKIDSS-UDS field. We first selected all $K$-band sources located within the ALMA primary beam (17$\farcs$4 FWHM) centred on the ALMA pointings, we then matched the $K$-band sources to the ALMA-detected SMGs presented in \citet{Simpson:2015ab} to within 1$''$ radius. By excluding one ALMA-detected SMG that is likely to be lensed by a nearby foreground source and therefore has its photometry contaminated (UDS286.0; \citealt{Simpson:2015aa}), in total the training sample comprises 164 $K$-band sources, of which 30 out of 52 ($\sim$60\% of the ALMA-detected SMGs) have $S_{850} > $\,1.5\,mJy. However, taking the primary beam correction into account our ALMA observations are only sensitive to sources with $S_{850} \geq $\,2.7\,mJy across the full ALMA primary beam. To make a clean comparison and to derive our model to separate SMGs from field galaxies, out of the 30 SMGs in the training sample, we only include the 22 that have S$_{850} \geq$\,2.7\,mJy as SMGs, and the rest with $S_{850} <$\,2.7\,mJy are regarded as part of the non-SMG comparison sample. 

One way to identify possible parameters that can be used to separate SMGs from non-SMG field galaxies is to search for correlations between $S_{850}$ and the chosen parameters. In order to do so we use the maximal information coefficient (MIC) statistics of the MINE package \citep{Reshef:2011aa} to determine the relative strength of the correlations between $S_{850}$ and the other source colors (for both ALMA-detected and -undetected sources), and to identify the primary criteria that isolate SMG counterparts from the contaminating field population located within the ALMA primary beam. The advantage of MIC over other correlation coefficient (such as Pearson and Spearman) is that it can identify non-linear relationship types , such as exponential or sinusoidal relation \citep{Reshef:2011aa}. 

%
%
\begin{figure}
\begin{center}
    \leavevmode
      \includegraphics[scale=0.85]{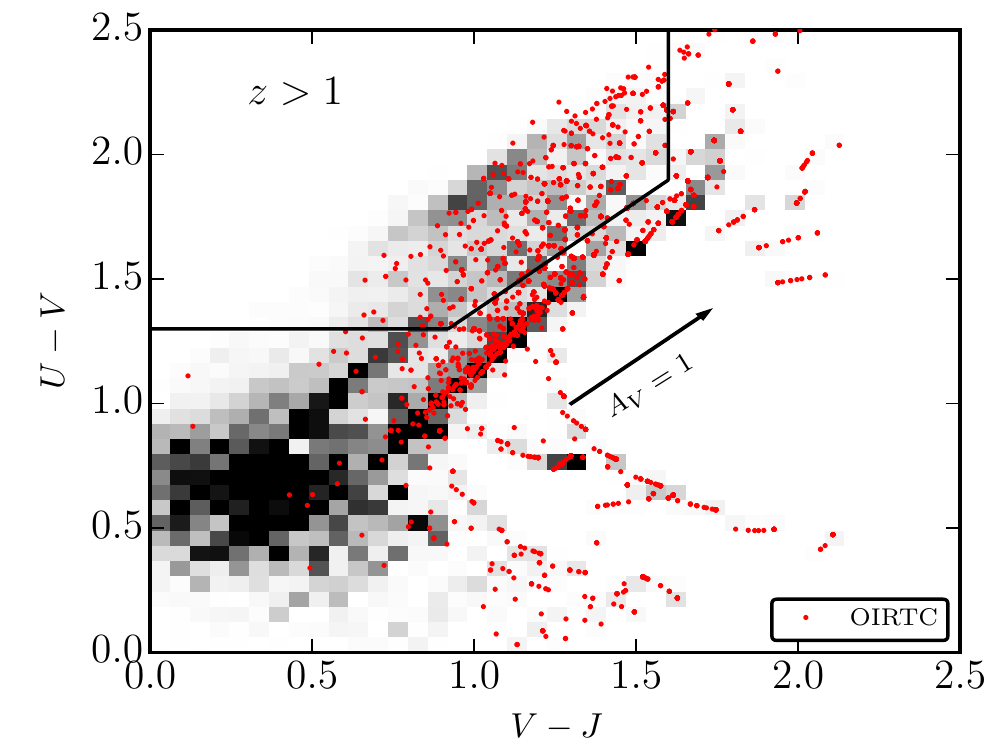}
       \caption{Rest-frame $UVJ$ diagram for $z>1$ sources. The red points represent the sources that are selected by the OIRTC technique, and the grey-scale background shows the density of the field galaxies in UDS, with higher density corresponding to darker color. 	Note that the pattern of the distribution is quantized due to the {\sc eazy} template fitting for deriving $z_{\rm photo}$ (\autoref{sec:ancillary}).
       This is to show that majority of the OIRTC-selected sources are at $z>1$ and occupies the color regions in which high dust extinction is expected. 
       }
    \label{uvj}
 \end{center}
\end{figure}

We select the following colors to search for correlations to $S_{850}$ (assigning zero flux to non-SMG comparisons): $(U-V)$, $(V-J)$, $(B-z)$, $(z-K)$, $(K-[3.6])$, $([3.6]-[4.5])$. Note that we only consider measurements that have at least 3\,$\sigma$ detections in both bands used in the color. 
The best correlation is found in the $(z-K)$ color, followed by $([3.6]-[4.5])$, $(K-[3.6])$, $(B-z)$, $(V-J)$, and $(U-V)$ \footnote[16]{We also note that the MIC analysis identifies radio emission as a powerful indicator of the correct ALMA counterpart, while indicating that MIPS 24\,$\mu$m emission provides no statistically compelling indication of the correct identification.}. In \autoref{colorhist} we compare the histogram of the SMGs and the non-SMG comparisons in each color. Indeed, the better the correlation in the MIC statistics, the better the separation between the two populations, as revealed in the higher fractions of SMGs in the redder colors, quantitatively expressed as the SMG fraction ($f_{\rm OIRTC}$). $f_{i, \rm OIRTC} = N_{i, \rm SMG}/N_i$ corresponds to the fraction of SMGs to the total number of sources in each color bin $i$. The errors of $f_{\rm OIRTC}$ are estimated through Monte Carlo simulations, in which we derive standard deviation of SMG fraction with 100 realisations of randomly populated data points based on their measured colors and errors, and the results are shown in the upper panels of \autoref{colorhist}. 

%
%
\begin{figure*}
\begin{center}
    \leavevmode
      \includegraphics[scale=0.8]{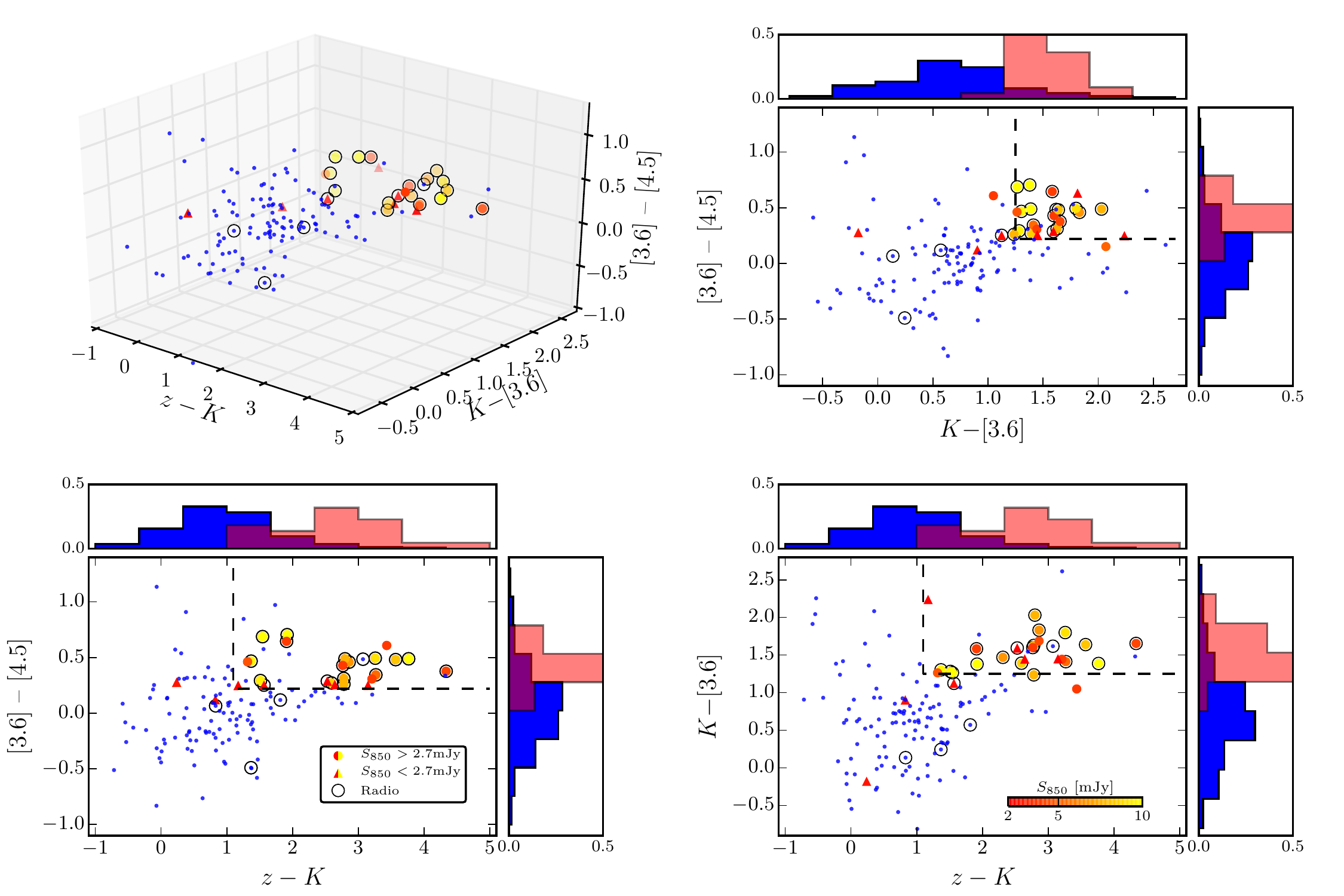}
       \caption{Triple-color ({\it top-left}) or color-color diagrams in $z-K$, $K-[3.6]$, and $[3.6]-[4.5]$. The large circles are SMGs, color-scaled based on their $S_{850}$, and blue dots are field sources. The top and left panels in the color-color diagrams are histrograms in each specified color, normalized to the total number of sources in each category, with assigned colors that are the same as those in \autoref{colorhist}. The SMGs are distinctively red compare to the non-SMG field sources in all three colors. The proposed color cuts (dashed lines) are given in \autoref{subsec:oir}.
       }
    \label{cccdiagram}
 \end{center}
\end{figure*}

In the three best correlated colors $(z-K)$, $(K-[3.6])$, and $([3.6]-[4.5])$, SMGs are mostly located in the redder part of the color space. At a typical SMG redshift, $z\sim$\,2, these colors correspond to roughly rest frame $(U-R)$, $(R-J)$, and $(J-H)$, suggesting both the Balmer/4000 \AA\, break and dust extinction could be the cause of SMGs being red in these colors \citep{Simpson:2014aa}. Indeed, in \autoref{uvj} we plot the rest-frame $UVJ$ color diagram of the UDS sample along with the sources that are selected based on these three OIR colors (OIRTC; described below), and find that OIRTC-selected sources are located in the regions where high $A_V$ is expected. Moreover, $>99$\% of the OIRTC-selected sources are located at $z>1$.

Next, motivated by the distinct red color space that SMGs occupy in $(z-K)$, $(K-[3.6])$, and $([3.6]-[4.5])$, we plot three-dimensional (3D) color-color-color along with two-dimensional (2D) color-color diagrams in \autoref{cccdiagram}. Interestingly, while the contamination fraction from the field sources is at best $\sim$30\% in the single color histograms (reddest bin in $z-K$; \autoref{colorhist}), the 2D and 3D color diagram efficiently remove most of the contaminants, revealing the red nature of SMGs as they are mostly clustered in the reddest color space. We note that the fainter SMGs with $S_{850} < 2.7$\,mJy have consistent colors to their brighter counterparts, except $[3.6]-[4.5]$, in which the fainter SMGs are bluer (median color 0.25$\pm$0.03 versus 0.46$\pm$0.04).

To select SMGs, by considering the product of accuracy and completeness, we propose a following triple-color cut

\begin{equation*}
(z-K) > 1.1 \wedge (K-[3.6]) > {1.25} \wedge ([3.6]-[4.5]) > 0.22
\end{equation*}

where $\wedge$ is the logical {\sc and} symbol. This triple-color cut works the best if the source has at least two color measurements. Based on the training sample, the triple-color cut successfully selects a SMG in 86$^{+14}_{-24}$\% (24/28) of the time and has a completeness of 46$\pm$11\% (24/52). 

While color cuts are easy to adopt, they do not consider information on the errors in the color measurements and the selected cuts are somewhat arbitrary. Below we employ a different approach, the OIRTC technique, that accounts for the uncertainties of the color measurements and define cuts in a quantitative way.

%
%
\begin{figure}
\begin{center}
    \leavevmode
      \includegraphics[scale=0.75]{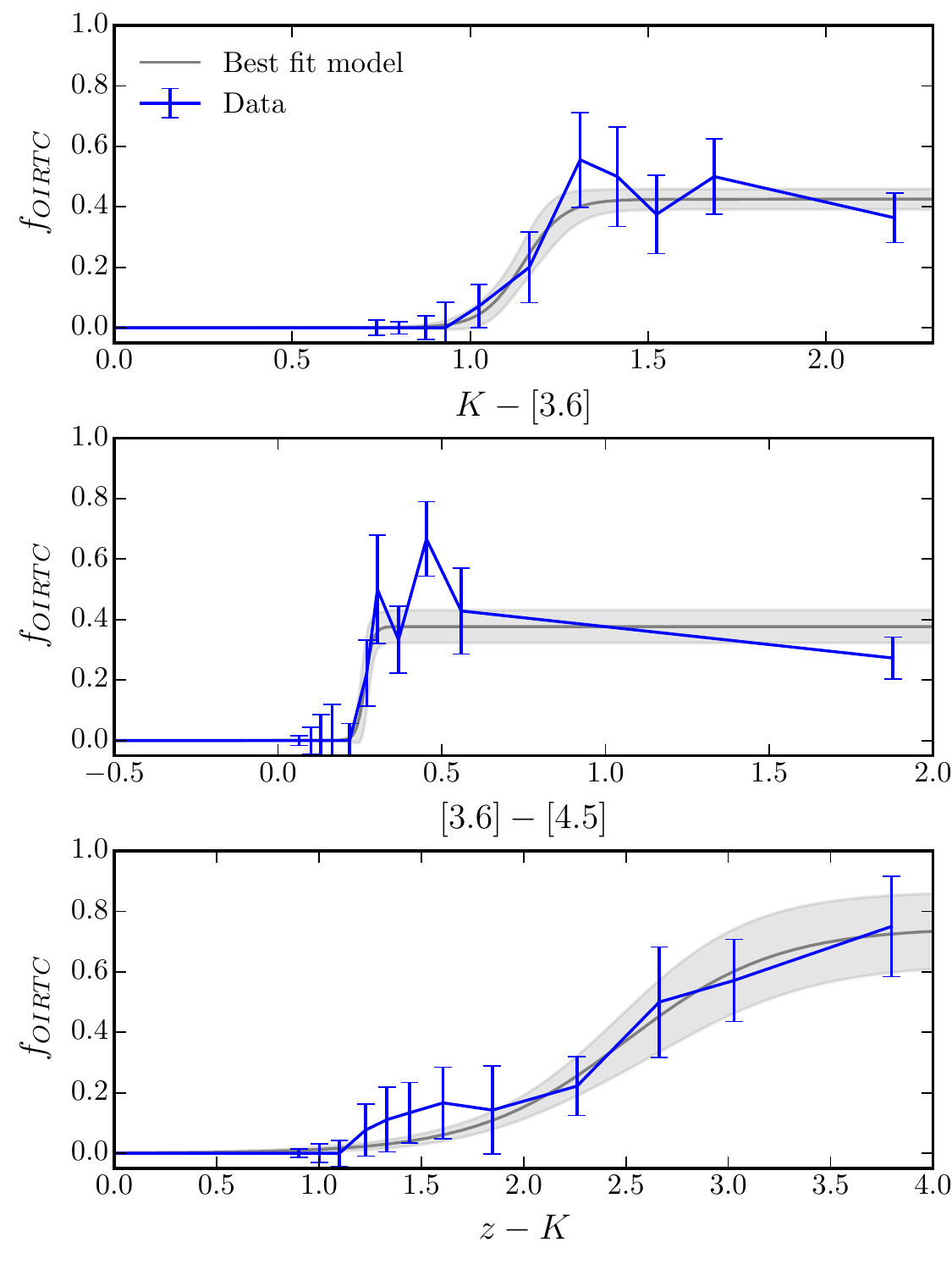}
       \caption{SMG fractions (${f}_{\rm OIRTC}$) for each specified color. The measurements are the same as those shown in \autoref{colorhist}, while the grey curves and the shaded regions are the best fit models with $\chi^2 \lesssim 1$ and their 1\,$\sigma$ errors. These models are then used to determine the cut that best separate the SMGs from the non-SMG field galaxies in the training sample. Detailed descriptions of the model fits are given in \autoref{subsec:oir}. 
       }
    \label{colormodel}
 \end{center}
\end{figure}

The SMG fraction ($f_{\rm OIRTC}$) shown in \autoref{colorhist} represents the fractional number density in each color bin, and its calculation takes color errors into account, as the uncertainties of $f_{\rm OIRTC}$ are obtained through Monte Carlo simulations. By describing $f_{\rm OIRTC}$ with model functions and calibrating the training set with those models, it is possible to determine cuts in a quantitative way. 

We therefore first model $f_{\rm OIRTC}$ as a function of color by fitting a functional form, parametrized as
$ 1/(a + e^{-b(x-c)}) $, in which $x$ is the corresponding color. The parametrisation is similar to the Fermi--Dirac distribution, which provides an appropriate description to the distribution of the measured $f_{\rm OIRTC}$, where the fraction on both sides of the color space converge, 
connecting by a smooth transition in between. When the colors are very red, the function converges to $1/a$, and to 0 when the colors are very blue. The parameter $b$ describes the sharpness of the transition whereas the parameter $c$ gives the color at which the value equals $1/(a+1)$. The best fit forms of this function are shown in \autoref{colormodel}, which all fit the distributions well ($\chi^2 \lesssim 1$), and we have confirmed that they are not sensitive to the chosen binning. 
The fitting results are given in \autoref{fit}

Based on the best-fit parameterised model, we then calculate the weighted-mean SMG fraction, $\langle f_{\rm OIRTC}\rangle$, defined as

\begin{equation}
\begin{split}
\langle f_{\rm OIRTC}\rangle = \frac{\sum_i f_{i, \rm OIRTC}\times W_i}{\sum_i W_i} \\
\end{split}
\end{equation}

where $i = z-K, K-[3.6], [3.6]-[4.5]$ and $W_i = N_i/\sigma_i^2$ represents the weight of each color, and $N_i = 148, 161$ and 147, respectively, is the number of available measurements in the training set for each color. Due to the sensitivity and coverage of the imaging, 80\% (132/164) of the training sample have all three colors measurements (i.e. $\geq$\,3\,$\sigma$ detections in both bands used in the color), and three field sources have only one or none. Thus the mean SMG fraction is weighted, for each color, by both the model uncertainties and the number of available measurements in the training sample. Because of the nature of this training sample, when applying the OIRTC technique to identify candidate counterparts in the whole SCUBA-2 sample, we only consider sources that have at least two color measurements. 

\begin{table}
\centering
\begin{threeparttable}
\caption{Best $\chi^2$ fits on the SMG fraction ($f_{\rm OIRTC}$)}
\begin{tabular}{cccc}
\toprule
 Color   &  a  &  b &  c   \\
\midrule
         $z-K$  &  1.34$\pm$0.31   & 2.71$\pm$0.80   &  2.61$\pm$0.24     \\
         $K-[3.6]$  &  2.35$\pm$0.31 & 17.40$\pm$12.74    &  {1.20}$\pm$0.08     \\
         $[3.6]-[4.5]$ & 2.65$\pm$0.34   & 80.0  & 0.28$\pm$0.02 \\
\bottomrule
\end{tabular}
 \begin{tablenotes}
      \small
      \item Note - All errors are obtained assuming $\Delta \chi^2 = 1$ except the b value for 
      $[3.6]-[4.5]$, which is the maximum value we set to prevent numerical overflow in the fitting
      process. Our results are not sensitive to this choice. 
    \end{tablenotes}
\label{fit}
\end{threeparttable}
\end{table}

\begin{figure}
\begin{center}
    \leavevmode
     \includegraphics[scale=0.47]{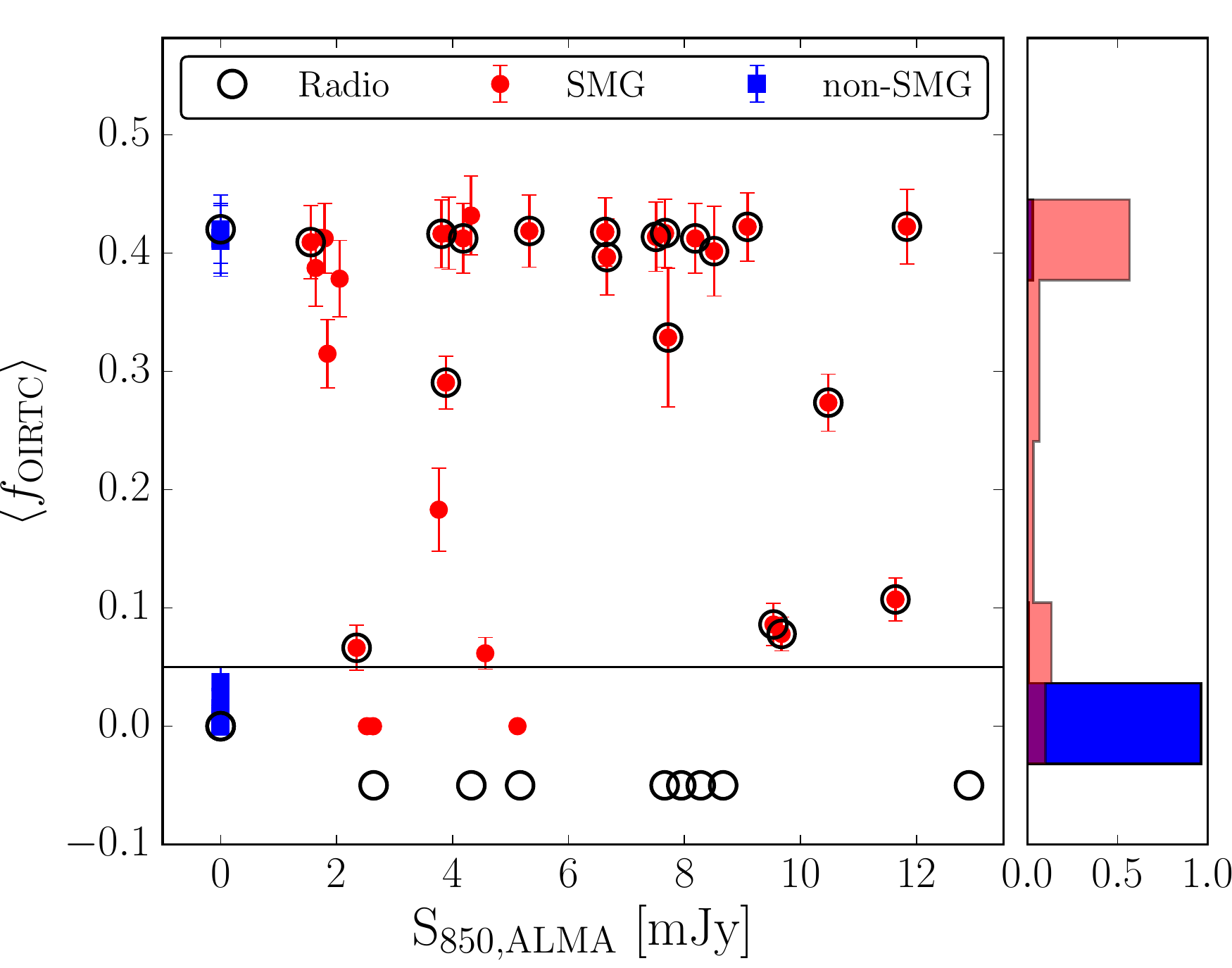}
       \caption{{\it Left}: The mean SMG fraction ($\langle f_{\rm OIRTC}\rangle$) of all 164 $K$-band sources that are located within the primary beam of our 30 ALMA observations in the UDS. For each source, based on the parameterised fits shown in \autoref{colormodel}, $\langle f_{\rm OIRTC}\rangle$ is calculated by taking the weighted averaging of all the corresponding values obtained from each measured color. We show both the non-SMG field sources (those not detected by ALMA) as well as the ALMA-detected SMGs. The color points enclosed by a circle are those with radio counterparts (matched within 1$\farcs$5). The empty black circles are radio-detected SMGs that do not have $K$-band counterparts, and thus we arbitrary set their $\langle f_{\rm OIRTC}\rangle$ to $-0.05$. The fact that most radio-detected, $K$-undetected ALMA SMGs have S$_{850} > 6$\,mJy and all $K$-detected, radio-undetected ALMA SMGs have S$_{850} < 6$\,mJy highlights the fact that these two ID methods compliment each other in flux space, which is further explored in \autoref{subsec:radpoirtc}. We find that a threshold of $\langle f_{\rm OIRTC}\rangle\geq$\,0.05, shown by the horizontal line, best separates the SMGs and the non-SMG field galaxies when judged on both accuracy and completeness. {\it Right}: The distribution of $\langle f_{\rm OIRTC}\rangle$ for SMGs and non-SMGs, normalised to the total number of sources in each category. Detailed discussions on this figure can be found in \autoref{subsec:oir}}
    \label{score}
 \end{center}
\end{figure}

Based on the best-fit models, we derive $\langle f_{\rm OIRTC}\rangle$ based on the color measurements of every source in the training sample, and we plot the results in \autoref{score}.
We find that, within the training sample, the mean SMG fraction of $\langle f_{\rm OIRTC}\rangle = 0.05$ best separates the SMG and non-SMG populations, if we maximise the product of accuracy and completeness.
Above the cut of the $\langle f_{\rm OIRTC}\rangle = 0.05$, the accuracy of correct SMG identification is 87$^{+13}_{-23}$\% (27/31), and the completeness is 52$\pm$12\% (27/52). Interestingly, although we only used SMGs with $S_{850} \geq$\,2.7\,mJy to derive models of the SMG fraction, the model is equally successful in identifying fainter SMGs, in which 75\% (6/8) are above the cut. While tentative evidence of bluer colors for SMGs with $S_{850} \lesssim 1$\,mJy has been reported by \citet{Hatsukade:2015aa}, we find that in the S$_{850} \gsim 1$\,mJy regime there is no strong color variations among SMGs with different fluxes. Furthermore, those non-SMG comparisons with high $\langle f_{\rm OIRTC}\rangle$ could also be faint SMGs with $S_{850} < $\,2.7\,mJy that are undetected by ALMA because they are located in the outskirts of the ALMA pointings where the sensitivity is slightly poorer. 

In summary, the OIRTC technique performs slightly better to the triple-color cut in both accuracy and completeness, although subject to the size of the training sample the differences are not statistically significant. We nevertheless adopt the OIRTC technique as the main method for selecting SMGs using optical-infrared colors as it performs the best in the training set. We note that our basic results are not sensitive to the chosen method.

\subsection{Our methodology: Radio+OIRTC identifications}\label{subsec:radpoirtc}

\begin{table}
	\centering
		\caption{Test results using the ALMA training sample}
		\begin{tabular}{ccc}
			\toprule
			Method   &  Accuracy  &  Completeness   \\
			\midrule
			Radio  &  87$^{+13}_{-23}$\% (26/30)   	 & 52$\pm$12\% (27/52)       \\
			MIPS  &  78$^{+22}_{-23}$\% (21/27) 	& 40$\pm$10\% (21/52)      \\
			OIRTC & 87$^{+13}_{-23}$\% (27/31)   	& 52$\pm$12\% (27/52)   \\
			Radio+OIRTC & 83$^{+17}_{-19}$\% (35/42)   	& 67$\pm$14\% (35/52)   \\
			\bottomrule
		\end{tabular}
\label{test}
\end{table}

The test results shown in \autoref{subsec:radio}, \autoref{subsec:mips}, and \autoref{subsec:oir} are summarized in \autoref{test}, which demonstrate that the accuracy of both radio and OIRTC identification is 87\%, while the MIPS identification is less accurate, with an additional issue of larger positional uncertainty. In addition, as hinted in \autoref{score}, the radio and OIRTC selection compliment each other in identifying SMGs in different flux ranges -- at S$_{850} >6$\,mJy five SMGs can only be identified by radio, while all the SMGs that can be selected by OIRTC technique, but are missed in radio, have S$_{850} < 6$\,mJy.

\begin{figure}
\begin{center}
    \leavevmode
      \includegraphics[scale=0.47]{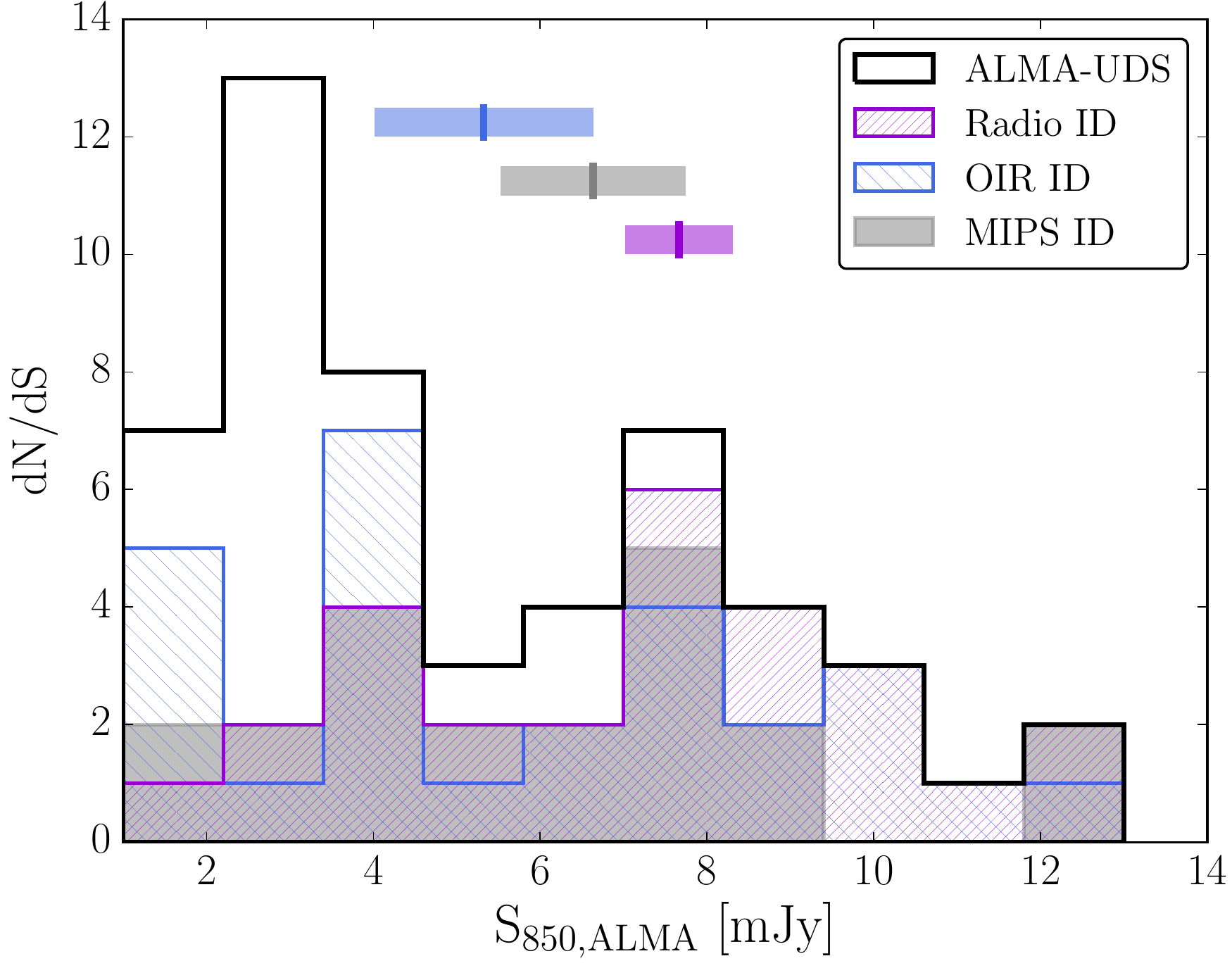}
       \caption{Histograms of $S_{850}$ for the 52 ALMA-detected SMGs in the UDS field from \citet{Simpson:2015ab}. The total sample is plotted, as well as
the various subsamples which are identified through the radio, optical-IR or MIPS mid-IR selection techniques (the vertical lines and corresponding horizontal bands show the median flux and the bootstrapped errors for
each subsample). We find that a given SMG can be identified by several different methods but the combination of radio+OIR finds most of the SMGs (\autoref{subsec:radpoirtc}).  
}
    \label{s850hist}
 \end{center}
\end{figure}

In \autoref{s850hist} we plot 850\,$\mu$m flux distribution of the ALMA SMGs, in which the sub-samples of SMGs identified by different methods are highlighted. Indeed, we find that while the radio sources preferentially identify brighter SMGs, the color analysis picks up fainter ones, and MIPS-identified SMGs have $S_{850}$ in between, as revealed in the median flux of SMGs each method identifies. By combining the radio sources and the OIRTC technique we are able to identify 35 out of 52 ALMA SMGs (67$\pm$14\% completeness), with the accuracy of 83$^{+17}_{-19}$\% (35/42). We emphasize that of all the 52 ALMA-detected SMGs, 14 have no counterpart in any of our ancillary images. In other words, the radio+OIRTC can identify all but three (92\%; 35/38) ALMA SMGs that are possible to be identified in other wavebands. We also find that all but one of the MIPS-identified SMGs can be selected through either radio or optical-IR color. In addition to the lower positional accuracy of MIPS sources, adding them into the counterpart selection method does not provide better results considering both accuracy and completeness. 

As a result of this analysis, in this study we adopt the radio+OIRTC technique as our major tool to identify the candidate SMG counterparts of SCUBA-2 detected submillimeter sources. Operationally this involves us taking all the radio sources that are matched to the submillimeter sources to within 8$\farcs$7, supplemented by the radio-undetected, $K$-selected sources identified using the OIRTC technique as candidate SMG counterparts. The chosen search radius of 8$\farcs$7, based on the theoretical calculations assuming Gaussian distribution ($\sigma \sim 0.6({\rm SNR})^{-1}$FWHM; \citealt{Ivison:2007fj}), corresponds to a 4\,$\sigma$ positional uncertainty for a 4\,$\sigma$ SCUBA-2 detection (14$\farcs$5 FWHM). The search radius of 8$\farcs$7 also matches to our ALMA primary beam, within which the training set is obtained.

%
%

\subsubsection{Testing our identification methodology}
We test our counterpart identification technique on two independent samples that are obtained from ALMA and SMA observations. The first sample of 12 is based on the SMA observations on 9 of our {\sc main} SCUBA-2 sources (none overlaps with the ALMA targets). The rms noise ranges between $\sigma = 1-2$\,mJy\,beam$^{-1}$, with a synthesized beam $\sim$2$''$ FWHM (Chapman et al.\ in prep.). The second test sample is the ALESS {\sc main} sample with 99 ALMA-detected SMGs, constructed by an ALMA follow-up study at 870\,$\mu$m on a flux-limited sample of 126 single-dish submillimeter sources detected on the LABOCA maps in the Extended {\it Chandra} Deep Field South (ECDFS; \citealt{Hodge:2013lr}). The ALESS observations have a median rms of $\sigma \sim $\,0.4\,mJy\,beam$^{-1}$, with a synthesized beam of $\sim$1$\farcs$6 FWHM. 

For the SMA sample, by excluding one source that is associated with a {\it Class} = 2 SCUBA-2 source (UDS.0010; although that source is accurately predicted by our identification method), we successfully identify 6 out of 11 sources (completeness = $55\pm28$\%) with an accuracy of $70^{+30}_{-34}$\% (7/10). While the SMA sample might be too small, both completeness and accuracy are consistent with our training results.

To compare to the second test sample, we first take the IRAC-based photometric catalog of sources in ECDF-S from \citet{Simpson:2014aa}. This includes 13-band photometry from the U-band to 8\,$\mu$m and derived photometric redshifts using {\sc hyperz} \citep{Bolzonella:2000aa} for $\sim$\,45,000 sources in the whole field. We then take those sources lying within the ALMA primary beam centred at the positions of the 88 LESS submillimeter sources from \citet{Weis:2009qy} for which there are good quality ALMA maps from \citet{Hodge:2013lr}. This yields a total of 326 sources and we match these to the ALESS main sample to within 1$\farcs$5 radius. This yields 64 ALESS SMGs with IRAC counterparts and 262 non-SMG sources. We also match to this catalogue the catalogue of 1.4\,GHz VLA $\geq $\,5\,$\sigma$ radio sources from \citet{Biggs:2011uq}. Finally we generate a counterpart candidate catalog based on our radio+OIRTC technique, in which we find an accuracy of {82$\pm$17\% (40/49)} and completeness of {40$\pm$8\% (40/99)}, which are consistent with the robust identifications based on the $p$-values presented in \citet{Biggs:2011uq} \citep{Hodge:2013lr}. 

This test result is very encouraging considering that the OIRTC model is derived based on our $K$-selected training sample in UDS, completely different than the IRAC-selected photometric sample in ECDF-S, yet our empirical method yields matched results to that based on the $p$-values. {Perhaps more importantly, as the ALESS-SMGs have fainter 850\,$\mu$m fluxes (median flux of $S_{850}$=2.5\,mJy versus $S_{850}$=4.2\,mJy for the ALMA-UDS sample), the equally high success rate for the ALESS sample suggests that the radio properties and/or the OIR colors do not vary significantly as a function of 850\,$\mu$m fluxes. This result reassures us that our method does not suffer from a bias owing to the fact that the ALMA training sample is biased toward brighter SCUBA-2 sources.}

A slightly lower completeness on ALESS, on the other hand, is caused by the fact that there are more fainter SMGs, which are more likely to have no detectable counterpart. {The finding of a different completeness in ALESS compare to that of our ALMA training sample also highlights the fact that the depth of the ancillary data affects the completeness of the identifications.}

\subsubsection{What are we missing?}\label{sec:miss}

\begin{figure}
\begin{center}
    \leavevmode
      \includegraphics[scale=0.47]{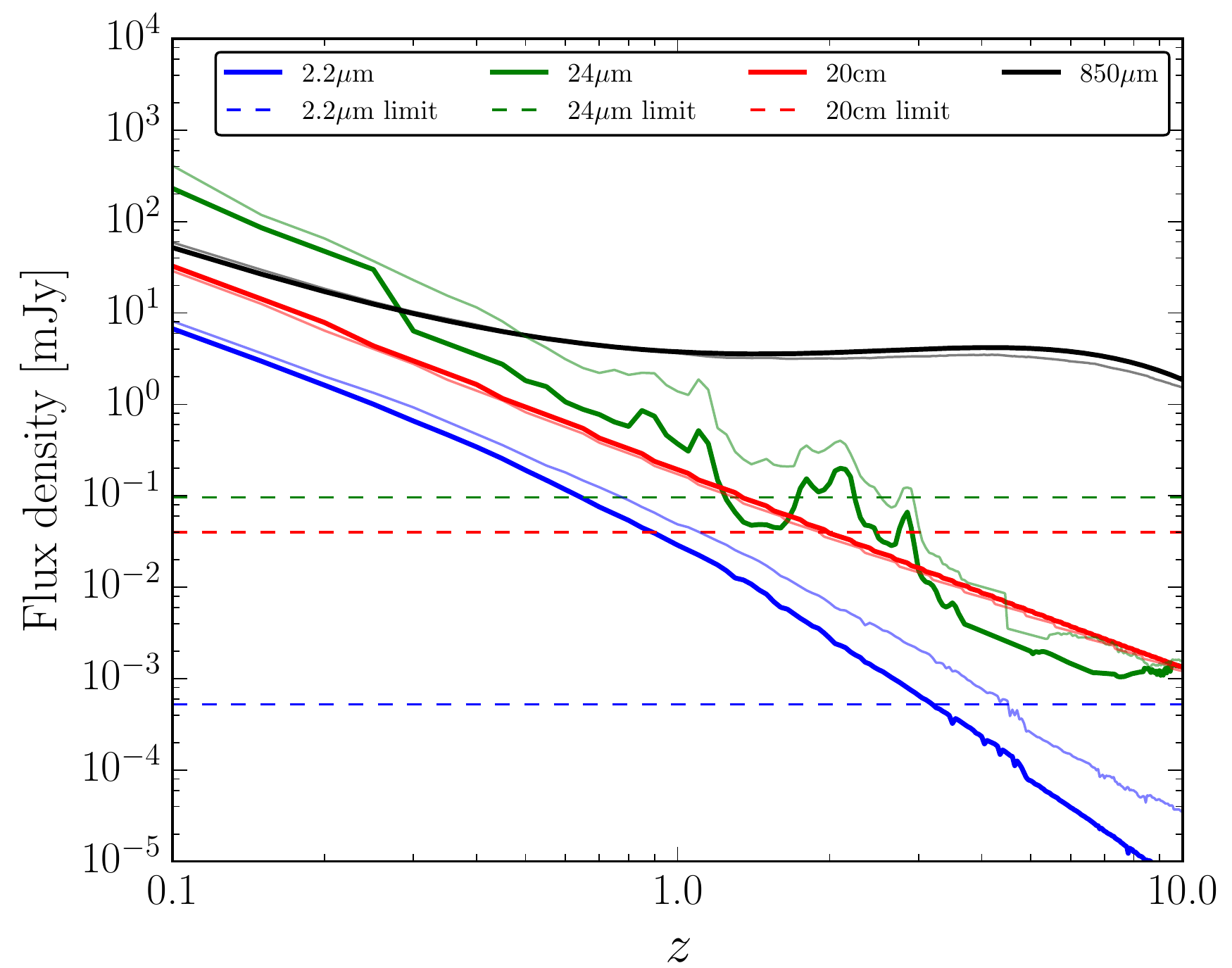}
       \caption{Expected flux densities in mJy as a function of redshift based on the SED of SMMJ2135-0102 (Cosmic Eyelash; \citealt{Swinbank:2010p10277}) normalised to a total SFR of 500\,M$_\odot$ yr$^{-1}$. The thin solid curves are redshifted flux densities based on the composite SED of the ALESS SMGs, with its infrared luminosity scaled to match that of the SMMJ2135-0102 track. Thick solid curves represent the predicted fluxes at 2.2\,$\mu$m, 24\,$\mu$m, 850\,$\mu$m and 20\,cm. The dashed horizontal lines marks the sensitivity of each of the identification waveband -- showing that the deep $K$-band imaging in UDS is sensitive enough to detect a given SMG to a highest redshift among the $K$/24\,$\mu$m/radio imaging.
 }
    \label{kcor}
 \end{center}
\end{figure}

Before we proceed and discuss the scientific implications, it is important to understand what SMGs are missed by our identification process. In \autoref{kcor} we plot the expected flux densities of an SMG as a function of redshift at 2.2\,$\mu$m, 24\,$\mu$m,  850\,$\mu$m and 20\,cm. We adopt an SED shape based on the Cosmic Eyelash, a strongly lensed SMG with a typical intrinsic $S_{850}$ similar to that of the SMGs we are probing \citep{Swinbank:2010p10277}, normalised to a total SFR of 500\,M$_\odot$yr$^{-1}$ (unlensed $S_{850}\sim3$\,mJy) assuming a Salpeter IMF. We also plot the results based on the composite rest-frame SED of the ALESS SMGs \citep{Simpson:2014aa, Swinbank:2014ul}. 

First of all, as the SEDs are matched to the same $L_{IR}$, it is not surprising that both SEDs have similar 850\,$\mu$m and radio fluxes as a function of redshift. However, ALESS SMGs have brighter observed 24\,$\mu$m fluxes and bluer color in optical/NIR as shown in a shallower decrease on the observed 2.2\,$\mu$m flux. We attribute this effect to the selection bias. As shown in \citet{Simpson:2014aa} and \citet{Swinbank:2014ul}, by necessity the composite rest-frame SED of the ALESS SMGs can only be constructed for those sources that have well-constrained photometric redshifts, meaning they are preferentially brighter, especially in the optical/UV. In addition, the comparable detection limits in the optical-IR wavebands in the ECDFS (the location of the ALESS SMGs) biases the detection towards bluer SMGs. 

\autoref{kcor} nicely illustrates that the negative $K$-correction of 850\,$\mu$m allows us to detect SMGs in a wide range of the redshift space \citep{Blain:2002p8120}, while optical-IR and radio suffer from positive $K$-corrections. The main implication is that our methodology of identifying counterpart candidates using radio/OIR imaging is likely to miss the high-redshift SMGs, as seen in \autoref{kcor} and many other studies in the literature (e.g., \citealt{Walter:2012qy, Riechers:2013yq}). We are also likely to miss more faint SMGs than the bright ones, providing that the mean redshift of the faint SMGs is similar or higher than that of the bright ones (e.g., \citealt{Chen:2014aa, Simpson:2014aa}), and the SEDs of the fainter SMGs do not differ significantly compared to the brighter ones. On the other hand, thanks to the deep $K$-band imaging available in UDS, given the same $L_{IR}$ higher-redshift SMGs are more likely to be detected in near-IR than in radio or at 24\,$\mu$m. Indeed, in \autoref{sec:z} we show that our OIRTC technique identifies SMG counterparts that have redshift distributions skewed towards higher-redshifts compared to those with radio counterparts.

%
%

\section{Results and Discussion}

\subsection{IDs for the entire SCUBA-2 UDS sample}\label{subsec:id}

We now apply our radio/optical-IR method to the full SCUBA-2 UDS sample. For the {\sc main} sample of 716 $\geq4\,\sigma$ submillimeter sources, we identify candidate counterparts to 498, from which 129 have two candidate SMG counterparts, 30 have three, and three submillimeter source have four SMG candidate counterparts. Example thumbnails on $\geq4\,\sigma$ submillimeter sources with $\ge$3 counterpart candidate are shown in \autoref{examplepostage}, and thumbnails for all the 1088 $\geq3.5\,\sigma$ source can be found  in the appendix.
Therefore we identify counterparts for 70$\pm$4\% (498/716) of the {\sc main} SMG sample, and the fraction of sources having multiple candidate counterparts (multiple fraction) is 33$\pm$3\% (163/498).

However, because not all the SCUBA-2 sources are covered by both radio and by more than three OIR wavebands needed for the OIRTC technique, each SCUBA-2 source is assigned to one of the three different classes, which are defined based on the following scheme.

1.	{\it Class} = 1: within the search radius of $8\farcs7$ from the SCUBA-2 position, sources that are covered by the radio imaging, and also qualified for the OIRTC technique (having at least two color measurements among $z-K$, $K-[3.6]$, $[3.6]-[4.5]$).

2.	{\it Class} = 2: sources that are only covered by the radio imaging, but lack the coverage necessary for the OIRTC technique (or only have it in part of the region within the search radius).

3.	{\it Class} = 3: sources that are not covered by the radio imaging, and not covered by the OIRTC technique either (or only covered in part of the region within the search radius).

For the 716 {\sc main} SCUBA-2 sources, 523 are $Class=1$, 191 are {\it Class} = 2 and 2 are {\it Class} = 3, respectively, and their spatial distribution is shown in \autoref{multicover}. For {\it Class} = 1 sources we identify candidate counterparts to 421, from which 124 have two candidate SMG counterparts, 28 have three, and three submillimeter source have four SMG candidate counterparts. 
Therefore for $Class =1$ {\sc main} sample we find an overall identification rate of 80$\pm$5\% (421/523) and a multiple fraction of 37$\pm$3\% (155/421). The numbers are much lower for {\it Class} = 2 {\sc main} sample, with an ID rate of $40\pm5$\% (77/191) and a multiple fraction of $9\pm4$\% (7/77). By construction there would be no ID for {\it Class} = 3 sources.

%
%
\begin{figure}
\begin{center}
    \leavevmode
      \includegraphics[scale=0.84]{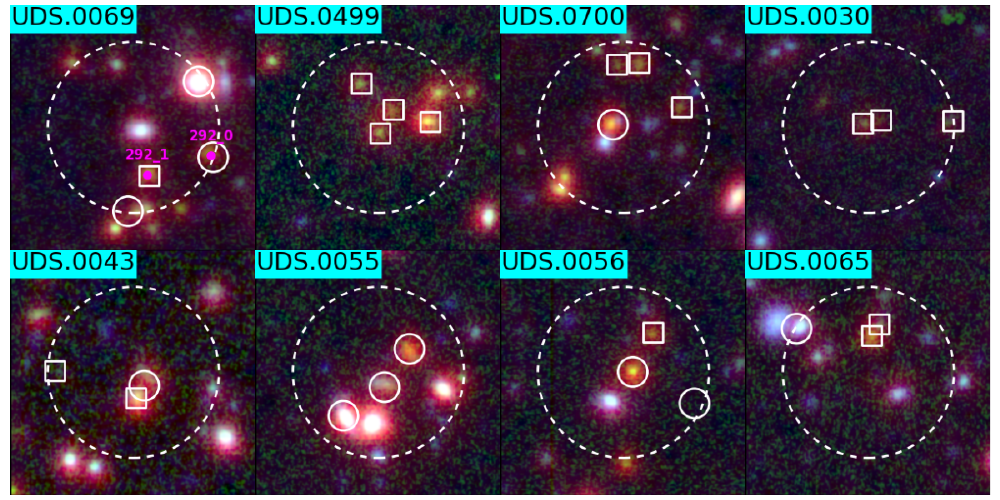}
       \caption{False-color [3.6]-$K$-$z$ (r-g-b) thumbnails of example $\geq4\,\sigma$ {\it Class} = 1 submillimeter sources with three or four candidate SMG counterparts. Each box is $25''\times25''$ and the large dashed circles show the counterpart searching area with a radius of 8$\farcs$7. The solid squares and circles mark the counterpart candidates identified through OIRTC technique and radio imaging, respectively. We only show circles if sources are identified by both radio and OIRTC. The magenta points are ALMA-detected SMGs with the ID numbers adopted from \citet{Simpson:2015ab}. All the example sources are {\it Class} = 1, which is shown with cyan background for each ID number (matched to the color scheme adopted in \autoref{multicover}). Detailed information on each counterpart candidate can be found in \autoref{tab:ids}, and the thumbnails of all the $\sim$ 1000 $\geq3.5\,\sigma$ submillimeter sources can be found in the Appendix.
 }
    \label{examplepostage}
 \end{center}
\end{figure}

Among the training sample, three of the ALMA pointings, including seven ALMA-detected SMGs, are classified as {\it Class} = 2 (UDS306, UDS47, and UDS408 in \citealt{Simpson:2015ab}), while the rest are {\it Class} = 1 sources. Therefore for the {\it Class} = 1 sources, based on the training sample we expect the identification accuracy to be 82$\pm$20\% (31/38), and the completeness to be 69$\pm$16\% (31/45), while for the {\it Class} = 2 sources where only radio coverage is available, based on \autoref{subsec:radio} we expect an identification accuracy of 87$^{+13}_{-23}$ and a completeness of 52$\pm$12\%. However, because the training sample was selected from the typically brighter ($S_{850} > 8$\,mJy) sources, the ALMA follow-up observations on the whole SCUBA-2 SMGs are likely to produce more fainter SMGs with $S_{850} < $\,4\,mJy than those in the training sample. We therefore expect the overall completeness to be less than the numbers quoted above. 

{To estimate the true completeness with our method on SMGs with $S_{850} > 1$\,mJy that are located within the beam area of the SCUBA-2 sources, we assume that we have ALMA follow-up observations for the rest of the SCUBA-2 sample, and we model the results by assuming for each SCUBA-2 source a 40\% chance that it breaks into two sub-components (see \autoref{subsec:multiplicity} regarding the multiple fraction for the SCUBA-2 sources). We then assume that for those broken into sub-components, the flux contribution is 75\% and 25\% of the SCUBA-2 flux, respectively, based on \citet{Simpson:2015ab}. We then apply the identification rate based on \autoref{s850hist} to the model flux distribution, and compute the completeness by dividing the expected number of identifications to the total number of the model SMGs. We obtain a completeness of $\sim$60\%.}

For the 258 {\it Class} = 1 submillimeter sources in the {\sc supplementary} 3.5--4\,$\sigma$ sample, we find candidate counterparts to 166, with 39 having multiple counterparts. The slightly lower fractions in identification rate (64$\pm$6\%) and multiple fraction (23$\pm$4\%) in the tentative sample could be due to the expected higher spurious rate ($\sim$\,10\%). The full identifications are given in \autoref{tab:ids}.

We stress that our counterpart identification method is trained based on the ALMA follow-up observations on part of our sample SCUBA-2 SMGs in UDS, and aims to provide all counterpart candidates with $S_{850} > 1$\,mJy.

\subsection{Identification rate and multiple fraction}\label{subsec:multiplicity}

%
%
\begin{figure}
\begin{center}
    \leavevmode
      \includegraphics[scale=0.47]{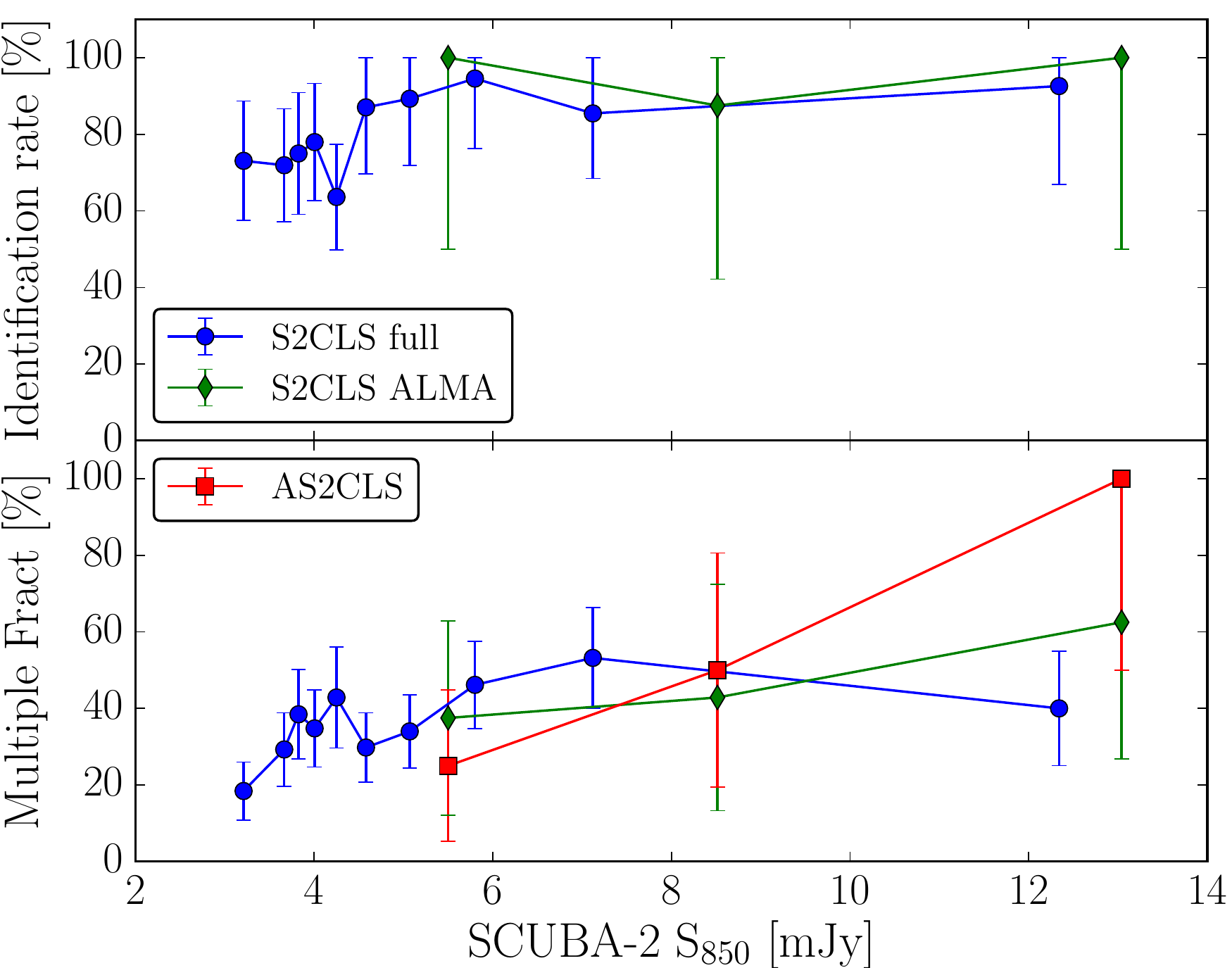}
       \caption{{\it Top}: The fraction of SCUBA-2 sources that have at least one candidate SMG counterpart (identification rate) in percentage as a function of measured $S_{850}$ for the submillimeter sources in the SCUBA-2 map. The errors are Poisson uncertainties for each equal-number bin. We show the results for the 524 $\geq4\,\sigma$ {\it Class} = 1 SCUBA-2 sources that are covered by radio and at least three of the $z$, $K$, 3.6\,$\mu$m or 4.5\,$\mu$m bands that are used for the OIRTC technique. We also illustrate the same quantities for those SCUBA-2 sources which were observed by ALMA in our Cycle 1 pilot study. We confirm a decreasing trend of identification fraction with decreasing $S_{850}$. {\it Bottom}: The fraction of SCUBA-2 sources that have more than one counterpart candidate (multiple fraction). The color coding is the same as the panel above, except that we also show the results based on the ALMA imaging on 27 of our SCUBA-2 sources. We argue that the true multiple fraction for single-dish submillimeter sources with $S_{850} \gsim $\,4\,mJy is likely $\geq40$\%, providing that follow-up observations are sensitive to $\sim$1\,mJy across the whole ALMA beam area.
       }
    \label{idrate}
 \end{center}
\end{figure}

In \autoref{subsec:radpoirtc}.1 we show that, for the $\geq4\,\sigma$ {\it Class} = 1 submillimeter sources, 81$\pm$5\% (the identification rate) have at least one counterpart candidate, and 37$\pm$3\% (multiple fraction) have more than one. In \autoref{idrate} we plot the identification rate and the multiple fraction as a function of the SCUBA-2 fluxes. We find a $\sim$90\% identification rate for submillimeter sources with $S_{850} > $\,5\,mJy, and $\sim$\,70\% for those with $S_{850}=$\,3--5\,mJy. 
The SCUBA-2 sources that were observed by ALMA are not significantly different. These results are in good agreement with ALESS \citep{Hodge:2013lr}, in which they also found $>$\,80\% ID rate for $S_{850} > $\,3\,mJy. The decrease in the identification rate for fainter submillimeter sources is likely to continue to the $S_{850} < $\,3\,mJy regime, which is shown to drop to $\lsim$\,50\% (e.g., \citealt{Hodge:2013lr, Chen:2014aa, Fujimoto:2015aa}). 

%
%
\begin{figure*}
\begin{center}
    \leavevmode
      \includegraphics[scale=0.8]{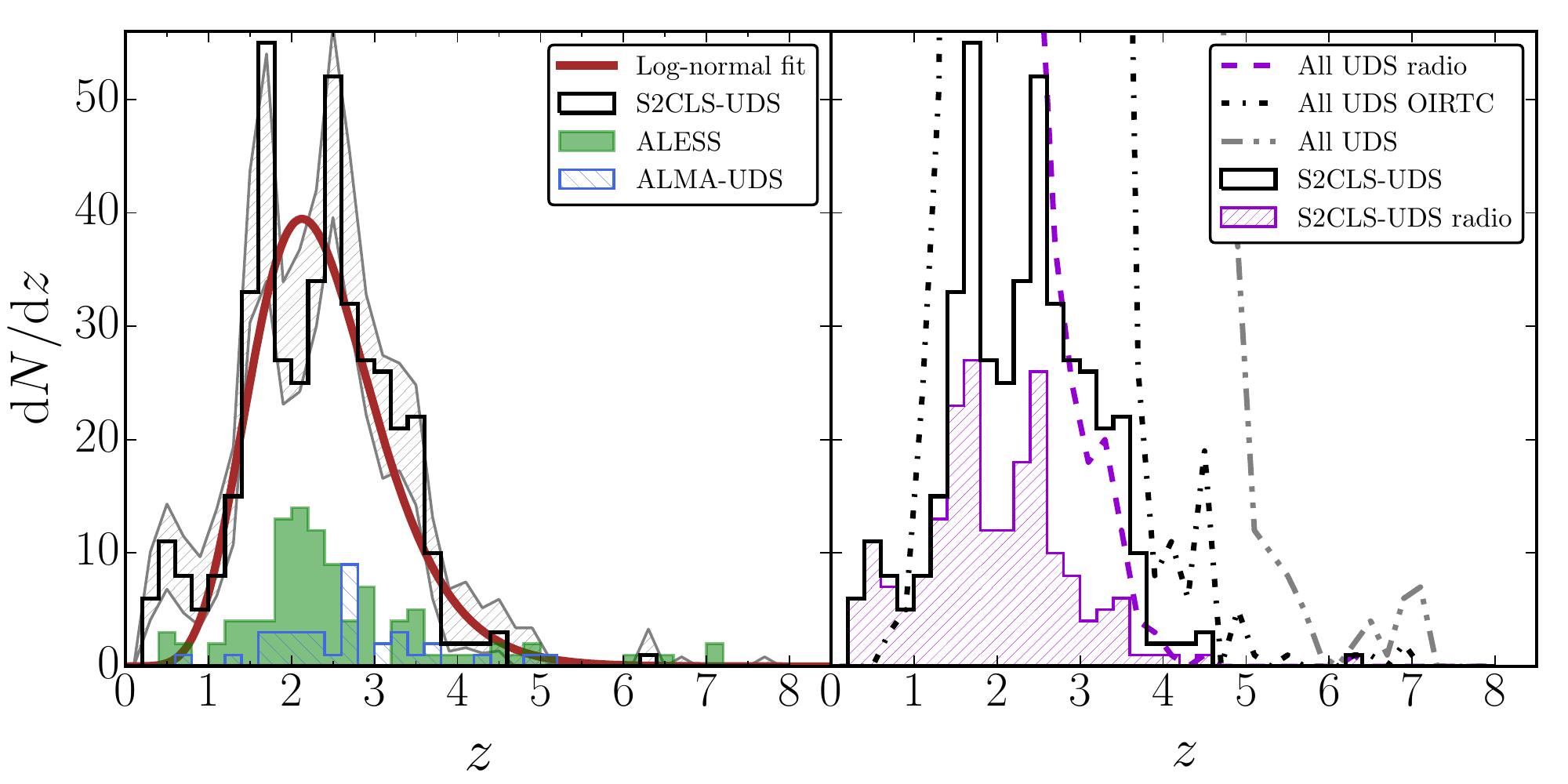}
       \caption{Redshift distributions of various samples. For both panels the black solid curves outline the distribution of all the candidate counterparts of the the $\geq4\,\sigma$ SCUBA-2 SMGs. {\it Left}: The grey hatched regions show the results of our Monte Carlo simulations, showing the uncertainties of the distribution by considering both the Poisson statistics and the errors of the photometric redshifts. The thick brown solid curve is the best-fit lognormal model to the grey hatched regions at $z<8$, with $\chi^2 = $\,1.4. The blue hatched regions show the redshift distribution of the ALMA-detected SMGs based on the ALMA follow-up observations of the 30 brighter SCUBA-2 sources in UDS (Simpson et al.\ in preparation). The green solid regions mark the distributions of the ALESS SMGs based on \citet{Simpson:2014aa}. {\it Right}: The purple hatched regions only show the candidate SMG counterparts that have radio detections. 
       To illustrate our selection biases, we also plot the distributions of all the radio sources, sources selected by the OIRTC technique, and the full UDS parent sample in dashed, dashed-dot, and dashed-dot-dot curves, respectively. 
       }
    \label{zhist}
 \end{center}
\end{figure*}

Interferometric follow-up observations have shown that single-dish SMGs are likely composed with multiple components. First tentative evidence for multiplicity came from radio-identified counterparts (e.g., \citealt{Ivison:2002uq}). Subsequently the true confirmation with submillimeter interferometry came from \citet{Wang:2011p9293} and many other studies \citep{Barger:2012lr, Barger:2014aa, Hodge:2013lr, Simpson:2015ab, Bussmann:2015aa}. However, these studies differ in the multiple fraction, which ranges from 10--70\% depending on the depth of the follow-up observations \citep{Chen:2013gq}. The most recent ALMA results suggest moderate multiplicity, $\sim$\,60\% for submillimeter sources with $S_{850}\sim$\,8\,mJy, with the primary source contributing on average $\sim$\,75\% of the the total flux from the single-dish source \citep{Simpson:2015ab}.  

In \autoref{idrate} we also show our results on the multiple fraction to the $\geq4\,\sigma$ {\it Class} = 1 SCUBA-2 sample. We find a constant multiple fraction of $\sim$\,40\% as a function of 850\,$\mu$m flux, with a positive (but not significant) slope. While the ALMA imaging reveals multiple fractions in broad agreement with ours, a noticeable but not significant increase (decrease) can be seen in the brightest (faintest) bin, making a correlation between multiple fraction and 850\,$\mu$m flux from SCUBA-2 slightly more significant. This can be explained by the fact that, in the multiple systems revealed by the ALMA imaging, many of the fainter companions have $S_{850} < $\,3\,mJy, making them unlikely to be identified through our method, or any other multiwavelength identification methods as they are usually undetected at almost all other wavelengths. On the other hand, a slightly lower multiple fraction in the faintest bin can be explained by the sensitivity of the ALMA imaging. The median sensitivity of the ALMA observations is 0.26\,mJy beam$^{-1}$, providing a 4\,$\sigma$ detection limit of $\sim$1\,mJy in the central region, and 2\,mJy within the primary beam. Assuming that the secondary sources contribute to 25\% of the total SCUBA-2 flux \citep{Simpson:2015ab}, the ALMA observations can detect the secondary SMGs for SCUBA-2 sources with $S_{850} > $\,8\,mJy. For SCUBA-2 sources fainter than $S_{850} < $\,8\,mJy, the ALMA observations are not sensitive enough to detect the secondary sources if located close to the edge of the primary beam, biasing the fraction towards a lower multiple value. We conclude that, for follow-up observations that are sensitive to $S_{850} \sim$1\,mJy across the whole ALMA beam area, the true multiple fraction for single-dish submillimeter sources with 850\,$\mu$m fluxes of $S_{850} \gsim $\,4\,mJy is likely to be higher than 40\%.

\subsection{Redshift distribution}\label{sec:z}

In \autoref{zhist} we plot the redshift distribution of the counterpart candidates of the $\geq4\,\sigma$ {\it Class} = 1 SCUBA-2 sources, in which we also show the distribution of those that have radio counterparts, ALESS \citep{Simpson:2014aa}, and our ALMA pilot study in UDS (Simpson et al.\ in preparation). 

The median redshift of the counterpart candidates of our $\geq4\,\sigma$ {\it Class} = 1 SCUBA-2 SMGs is $z=2.3\pm0.1$, and that based on the radio identifications is $z=1.9\pm0.1$. The median redshift of the ALMA-UDS sample is slightly higher at $z=2.7\pm0.2$. The difference is not significant but this could suggest a dependency between 850\,$\mu$m flux and redshift, since the ALMA-UDS sample is much brighter. By conducting photometric redshift analysis on ALESS SMGs, \citet{Simpson:2014aa} found a weak trend between 870\,$\mu$m flux and redshift. However, after accounting for the selection bias Simpson et al. conclude that the median redshift is likely not dependent on the 870\,$\mu$m flux. A weak positive or non-existing trend is in contrast with recent predictions by \citet{Cowley:2015aa}, who based on their semi-analytic model predict a negative trend between 850\,$\mu$m flux and redshift. 

To obtain errors of the redshift distribution, we perform Monte Carlo simulations based on both the Poisson statistics and uncertainties in the photometric redshifts. We take the errors estimated from the Monte Carlo simulations and fit the distribution with a lognormal function described as

\begin{equation}
\frac{{\rm d}N}{{\rm d}z}(z) = \frac{A}{\sqrt{2\pi}(1+z)B}\mathrm{e}^{-\frac{[ln(1+z)-ln(1+z_\mu)]^2}{2B^2}}.
\end{equation}
The best-fit parameters are $A=74.6\pm4.3$, $B=0.23\pm0.01$, and $z_\mu=2.30\pm0.05$, with a reduced $\chi^2=1.4$.

In addition, we observe two peaks in our redshift distribution, one at $z\sim$\,1.6, and the other at $z\sim$\,2.5 (although considering both the Poisson statistics and uncertainties in the photometric redshifts, a lognormal formalism still provides good fit to the redshift distribution). The former corresponds to the known $z=$\,1.62
galaxy cluster  Cl\,0218.3$-$0510 (e.g., \citealt{Papovich:2010aa}) which our previous work has shown likely contains a population of 
submillimeter-detected ULIRGs \citep{Smail:2014aa}. In contrast, there is no pre-known large scale structures at $z\sim$\,2.5 in UDS.

\begin{figure}
\begin{center}
    \leavevmode
      \includegraphics[scale=0.8]{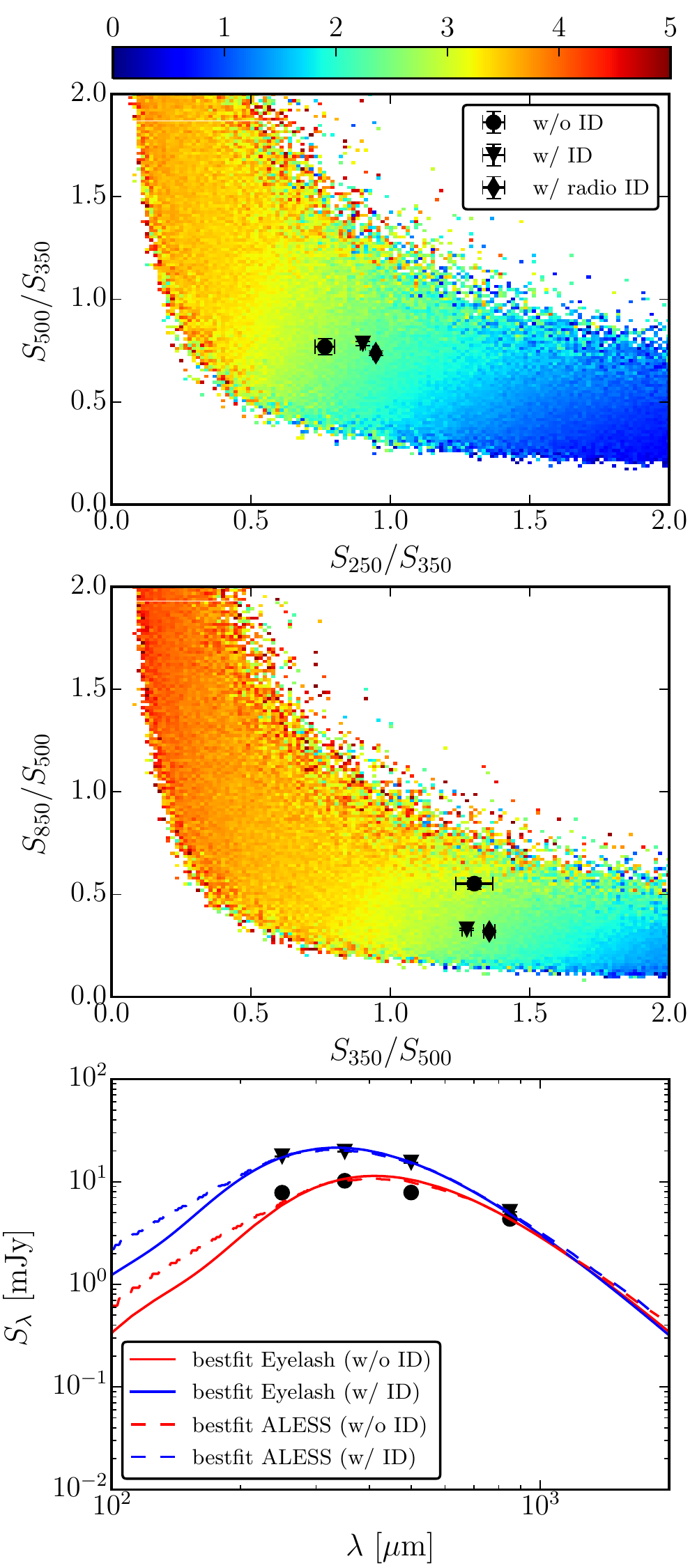}
       \caption{{\it Top} and {\it Middle}: FIR color-color diagram of SCUBA-2 sources stacked on SPIRE/SCUBA-2 maps. The SCUBA-2 sources that have (radio) counterpart candidates are shown in {\it downward triangle} ({\it diamond}), and those that do not have any counterpart identification is plotted as a {\it circle}. Most errors are smaller than the symbols. The other points color-coded by redshifts, are the expected FIR colors assuming single temperature grey-body spectra, in ranges of $15\leq T_d\leq45$, $1\leq \beta\leq2.5$, and $0\leq z\leq5$. The redder color of SCUBA-2 sources without any counterpart identification suggests that these sources are likely to be at higher redshift than those that have candidate counterparts. {\it Bottom}: stacked SEDs of the SCUBA-2 sources that do and do not have counterpart identification, along with the best-fit template SEDs based on SMMJ2135-0102 and ALESS SMGs. The best-fit redshift of the SCUBA-2 sources with(without) counterpart identifications is 2.5(3.3), consistent with the higher-$z$ nature for sources without candidate counterpart suggested by the FIR color-color diagram.}
    \label{fircolor}
 \end{center}
\end{figure}

As discussed earlier, our redshift distribution is likely to be biased against high-redshift sources, which are essentially too faint to be detected in the NIR survey (\autoref{subsec:radpoirtc}.2). To illustrate this bias in another way, we measure the weighted average fluxes at 250, 350, 500, and 850\,$\mu$m by stacking on the SPIRE and SCUBA-2 maps at the SCUBA-2 source positions. The measurements are made on the SCUBA-2 sources that have candidate SMG counterparts, those that have radio candidate counterparts, and those without any counterpart identification. We plot the results as FIR colors in \autoref{fircolor}, in which we find that SCUBA-2 sources with radio identifications have the bluest color, followed by those with any kind of identifications, and those without counterpart candidate are the reddest. Following \citet{Amblard:2010aa} and \citet{Ivison:2012aa}, we derive expected FIR colors on $10^{6}$ grey-body spectra assuming a single dust temperature, $T_d$, where the flux density $f_\nu \propto \nu^{3+\beta}/(exp(\frac{h\nu}{kT_d})-1)$. To account for the flux uncertainties in the data, each model flux density is randomly deviated by 10\% assuming Gaussian distribution. We explore the parameter space at $15\leq T_d\leq45$, $1\leq \beta\leq2.5$, and $0\leq z\leq5$, and the results are plotted in \autoref{fircolor} in color scale. Clearly, the redder the FIR color, the higher redshift the source is likely to be at. 

In \autoref{fircolor} we also plot the stacked SEDs along with the best-fit template SEDs based on SMMJ2135-0102 and ALESS SMGs. We find the with either template the best-fit redshift of the SCUBA-2 sources that have candidate counterparts is $z=2.5$, while that of the sources without any candidate counterpart is $z=3.3$, consistent with the higher-$z$ nature on sources without candidate counterpart suggested by the FIR color-color diagram. 

The redder FIR color of the SCUBA-2 sources without counterpart identification supports the idea that these sources are on average at higher redshift compare to those with counterpart candidates. Furthermore, the fact that sources with both radio and OIRTC identifications are slightly redder those that only have radio counterparts also supports the idea stated in the previous sections that OIRTC technique finds candidates at slightly higher redshifts.

To roughly estimate the median redshift of the complete $\geq4\,\sigma$ {\it Class} = 1 submillimeter sources, we set the redshifts for that 20\% (102/523) that do not have any counterpart candidate to $z=3$. We assume a multiple fraction of 40\% based on the results from \autoref{subsec:multiplicity}, and we find the median redshift shifting to $z=2.6\pm0.1$. This result is not sensitive to the assumed multiple fraction.

%
%

Recently, simulations have suggested that redshift distributions of dusty galaxies are dependent to their selecting FIR wavelength, in a sense that observations at longer wavelengths tend to select higher redshift sources (e.g., \citealt{Zavala:2014aa, Bethermin:2015aa}). This selection bias is also supported by many observations (e.g., \citealt{Chapin:2009aa, Smolcic:2012pp, Yun:2012aa, Casey:2013aa, Weis:2013ly, Miettinen:2015aa}). While in some cases the differences could be negligible considering the uncertainties of the measurements, to ensure like-to-like comparisons, we only compare our results to those that are also based on the observations that were carried out at 850/870\,$\mu$m. 

\citet{Chapman:2005p5778} (C05) reported a median redshift of $2.2\pm0.1$ based on a sample of SCUBA-detected, radio-identified, and spectroscopically confirmed SMGs. Our radio-identified counterpart candidates have a median redshift $z=1.9\pm0.1$, slightly lower but still consistent with that of C05. This is expected as the radio imaging used in C05 is on average slightly deeper than that used in this work, and thus C05 might select SMGs with slightly higher redshifts (\autoref{subsec:radpoirtc}.2). Indeed, by only considering samples of C05 that are covered by 1.4\,GHz VLA imaging with a radio depth of $1\sigma \geq 10$\,$\mu$Jy, the median redshift of C05 sample becomes $z=1.9\pm0.4$.

\citet{Simpson:2014aa} reported a median redshift of $z=2.3\pm0.1$ for 77 ALESS ALMA SMGs that have sufficient optical and NIR photometry to derive reliable photometric redshifts. Our result is in excellent agreement with that of \citet{Simpson:2014aa}. Furthermore, for the remaining 19 ALESS SMGs that do not have sufficient photometry, \citet{Simpson:2014aa} argued that these are likely at $z>$\,3 and by placing them in the high-redshift tail the median redshift is raised to $z=2.5\pm0.2$, which is again in good agreement with our estimate after accounting for the SCUBA-2 sources that do not have any identified counterpart. 

On the theoretical front, \citet{Zavala:2014aa} predicted the median redshift to be $z=2.43\pm0.12$, and \citet{Cowley:2015aa} showed that for sources with $S_{850} > 1$\,mJy, which is the flux regime probed by our training sample, the median redshift is $z=2.77\pm0.11$.These are all again consistent with our results.

\subsection{SMG clustering}\label{sec:clustering}
The significant improvement of the SMG sample size in degree-scale fields provides an unique opportunity to investigate the clustering properties of the SMGs. To study the SMG clustering, we calculate the two-point autocorrelation function $w(\theta)$ using the Landy \& Szalay (1993) estimator.

\begin{equation}\label{eqn:ls}
w(\theta) = \frac{1}{RR}(DD-2DR+RR)
\end{equation}
where $DD$, $DR$, and $RR$ are the number of Data-Data, Data-Random, and Random-Random galaxy pair, respectively, counted in bins of separations $\theta$. $DR$ and $RR$ are normalized to have the same total pairs as $DD$, in a sense that given $N_{\rm SMG}$ SMGs, $N_R$ random points, $N_{gr}(\theta)$ and $N_{rr}(\theta)$ in the original counts, $DR = [(N_{\rm SMG}-1)/2N_R]N_{gr}(\theta)$ and $RR = [N_{\rm SMG}(N_{\rm SMG}-1)]/N_R(N_R-1)]N_{rr}(\theta)$.

Because our sample of SMGs are located in a single, area-limited region, $w(\theta)$ could be biased due to the fact that the mean density measured in our data is not the true underlying mean density over the whole sky. The mean density is usually biased high, making the observed clustering appear weaker than the true value. If the real SMG $w(\theta)$ can be described as a power-law model $w(\theta)_{true} = A\theta^{-0.8}$ (which were found to be true both observationally and theoretically, at the physical separation of $\sim$0.1--10 $h^{-1}$ Mpc), the observed $w(\theta)$ will follow the form 
\begin{equation}\label{eqn:wth}
w(\theta) = w(\theta)_{true} - IC, 
\end{equation}
with the bias $IC$ known as the integral constraint. The integral constraint can be numerically estimated (e.g., \citealt{Infante:1994aa, Adelberger:2005aa}), using the random-random pairs under the form 
\begin{equation}
IC=\frac{\sum_i N_{rr}(\theta_i)w(\theta_i)_{true}}{\sum_i N_{rr}(\theta_i)}.
\end{equation}

In practice, in \autoref{eqn:ls} we use 4 times as many random points as the number of SMGs (data points) and repeat the estimate 25 times. Using these 25 estimates we calculate the variance, the mean $w(\theta)$, as well as the mean $N_{rr}$ for the correction of the integral constraint. We then perform $\chi^2$ minimization using \autoref{eqn:wth} to find the best fit $w(\theta)_{true}$ on $0\farcm2-6'$ scale ($\sim0.2-6 h^{-1}$Mpc at $z=2$), the power-law regime that is shown below. At this stage the error of the amplitude A in $w(\theta)_{true}$ is unrealistically small as the variance only accounts for the shot noise from the the creation of the random points and the Poission uncertainties of the $DD$ counts ($DD^{0.5}$). 

To estimate the systematic uncertainties due to field-to-field variation we conduct the ``delete one jackknife'' resampling method \citep{Norberg:2009aa}. We first divide the chosen rectangular area\footnote[17]{For consistency and the ease of estimating the jackknife uncertainties, we only use counterpart candidates for the {\it Class} = 1 submillimeter sources that are located within a chosen rectangle region with a size of $\sim$0.5 degree$^2$ ($0.65^{\circ}\times0.78^{\circ}$).
}, in which we calculate $w(\theta)$ for the whole sample, into $N_{sub} = 9$ ($3\times$3) equal-size sub-area. Each jackknife sample is defined by discarding, in turn, each of the $N_{sub}$ sub-area into which the whole sample has been split. Each jackknife sample therefore consists $N_{sub}-1$ remaining sub-area, with a volume ($N_{sub}-1$)/$N_{sub}$ times the volume of the full rectangular area. The $w(\theta)_{true,jk}$ fit is repeated, in each jackknife sample, for $N_{sub}$ times (as there are only $N_{sub}$ jackknife sample by construction) based on the method described in the previous paragraph. The systematic uncertainties are then estimated through the variance of these $w(\theta)_{true,jk}$ fits.
Finally, the error of the amplitude $A$ for $w(\theta)_{true}$ is computed by accounting for the shot noise, Poission errors, and the systematic uncertainties estimated based on the jackknife resampling. 

%
%
\begin{figure}
\begin{center}
    \leavevmode
      \includegraphics[scale=0.88]{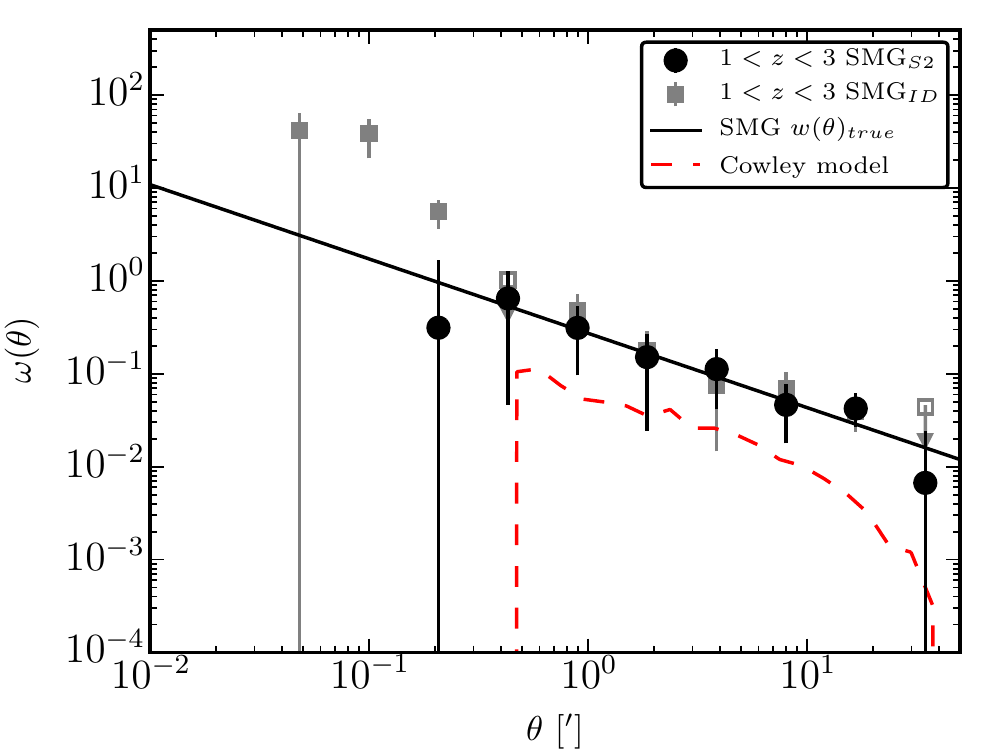}
       \caption{The angular autocorrelation function of $1<z<3$ SMGs ({\it black circles} and {\it grey squares}),
       all corrected for the integral constraint. While both are measured from the SCUBA-2 detections that have counterpart candidates at $1<z<3$, SMG$_{S2}$ represents measurements based on the positions of the SCUBA-2 detections, and those marked as SMG$_{ID}$ are based on the positions of the counterpart candidates. Both measurements are consistent at $>0\farcm3$ (about the size of the ALMA beam) scales, but those of SMG$_{ID}$ are artificially enhanced at $<0\farcm3$ scales due to selection biases (see \autoref{sec:clustering} for detail discussions). The uncertainties shown for each datapoint only include shot noise and Poission errors. The best-fit power-law model for the SMGs 
       is shown as {\it solid black}, while the predictions for the SCUBA-2 surveys by \citet{Cowley:2015ab} are plotted as {\it dashed red} curves. 
       We detect strong clustering signals on SCUBA-2 detected sources, higher than but statistically consistent with the predictions of \citet{Cowley:2015ab}. 
       }
    \label{wtheta}
 \end{center}
\end{figure}

In \autoref{wtheta} we show the $w(\theta)$ measurements (corrected for the integral constraint) of the 169 {\it Class} = 1 {\sc main} submillimeter sources that have candidate SMG counterparts at $1<z<3$, the redshift range where most of the counterpart SMGs lie and our selection is likely to have higher completeness. While we use the probability distribution of redshifts, $p(z)$, of the candidate SMGs, we use the positions of the SCUBA-2 sources, not the positions of the candidate SMGs. This is because while the SCUBA-2 sources are selected uniformly with $S_{850}\gtrsim 3$\,mJy, the candidate SMGs might have fluxes down to $S_{850}\sim$1\,mJy as they are based on a training set from the deeper ALMA imaging. By using all candidate SMGs, we are effectively creating a sample which is subject to uneven flux selections within and outside the beam, potentially resulting in an artificial boost in $w(\theta)$ on small scale (\autoref{wtheta}). This is not to say that the clustering of SMGs is sensitive to 850\,$\mu$m flux, in fact as we show in \autoref{wtheta} the results at $>0\farcm3$ (greater than the size of the beam) are not sensitive to the adopted positions. At this scale the clustering is determined by the original SCUBA-2 catalog. 

Our clustering measurement is a factor of four higher, though still within uncertainties, than the theoretical predictions for the submillimeter single-dish surveys recently proposed by \citet{Cowley:2015ab}. Our results are also higher, but still within uncertainties, than the measurements reported previously \citep{Webb:2003p6591, Blain:2004aa, Weis:2009qy, Williams:2011aa, Hickox:2012kk}.

Based on the dark matter (DM) power spectrum provided by the {\sc halofit} code of \citet{Smith:2003aa}, our $w(\theta)$ measurements suggest a galaxy bias of $b = 9.1^{+2.1}_{-2.8}$ and an autocorrelation length of $r_0 = 21^{+6}_{-7} h^{-1}$Mpc, which corresponds to a DM halo mass of $M_{halo} = 8\pm5 \times10^{13} h^{-1} M_\odot$.  While most results report SMG DM halo mass of $M_{halo} \sim 10^{13} h^{-1} M_\odot$, it is apparent that the measurements of SMG clustering are still suffer from large uncertainties. Larger samples with better determinations of their redshifts, or cross-correlation with other populations with larger sample size (Wilkinson et al. in prep.), are needed to provide better constraints to the SMG clustering properties.

\section{Summary}
We present the results of the identification of counterparts to 1088 submillimeter sources that are detected at $\geq3.5\,\sigma$ in our SCUBA-2 Cosmology Legacy Survey imaging of the UKIDSS-UDS field. We analyse a subset of 716 $\geq4\,\sigma$  SCUBA-2 sources, expected to have high fidelity with $\sim1$\% false detection rate. The SCUBA-2 sources are categorised into three classes, based on their multi-wavelength coverage, with the {\it Class} = 1 sources having the best coverage for the counterpart search. The analyses of this paper is built upon an ALMA pilot study on a subset of 30 brighter SCUBA-2 sources \citep{Simpson:2015ab,Simpson:2015aa}, as well as lessons learnt from the ALESS ALMA survey of submililmeter sources in the ECDFS \citep{Hodge:2013lr}, and the results are summarized as following: 

\begin{enumerate}
\item Based on ALMA observations of a subset of SCUBA-2 sources, we investigate the accuracy and the completeness of the $p$-values that has been widely applied in the literature to find radio/MIPS counterparts for the submillimeter sources. We find that at the depth of our ALMA imaging (central $1\,\sigma\sim0.25$\,mJy beam$^{-1}$), the accuracy of both radio and MIPS identifications is not dependant on the $p$-value for $p < 0.1$, {although MIPS has a poorer spatial resolution, leading to more blending}. We find that including all the radio and MIPS sources that are located within the ALMA primary beam produces better identification results, in terms of maximising the product of accuracy and completeness, compare to the traditional method of only considering the $p<0.05$ sources. By doing so we find the accuracy and the completeness of the radio (MIPS) identification to be $87^{+13}_{-23}$\% ($78^{+22}_{-23}$)\% and $52\pm12$\% ($40\pm10$\%).
 
\item Using our 52 ALMA-detected SMGs in the UDS as a training set, we develop a novel technique for counterpart identification, OIRTC, by utilising three optical and NIR colors ($z-K$, $K-[3.6]$, and $[3.6]-[4.5]$). For sources above the mean SMG fraction cut $\langle f_{\rm OIRTC}\rangle > 0.05$, the OIRTC technique provides accuracy and completeness almost identical to the radio identification, $87\pm23$\%, and $52\pm12$\%, respectively. Most importantly, the OIRTC technique complements the radio identifications in selecting SCUBA-2 sources, in a sense that OIRTC selects fainter sources. In addition, the OIRTC technique recovers almost all MIPS identifications. Based on these results, we adopt both radio imaging and the OIRTC technique (radio+OIRTC) to select counterpart candidates in this work. 

\item In the two OIR colors that are used to train the OIRTC technique, $z-K$ and $K-[3.6]$, we find in the $S_{850} \gtrsim1$\,mJy regime that there is no strong color variations among SMGs with different 850\,$\mu$m fluxes. In $[3.6]-[4.5]$ color, however, we find that the fainter SMGs with $S_{850} < 2.7$\,mJy have a median color bluer than that of the brighter SMGs ($S_{850} > 2.7$\,mJy). 

\item For the 523 $\geq4\,\sigma$ {\it Class} = 1 SCUBA-2 sources that have both radio and OIRTC coverage, we find at least one candidate counterpart for $80\pm5$\% of the sample, and $37\pm3$\% have more than one candidate counterpart. Based on the training sample, the identifications of this sample is accurate to $82\pm20$\%, with a completeness of $69\pm16$\%, although the completeness may be lower ($\sim$60\%) due to the fact that the training sample is based on bright SCUBA-2 sources. {The fact that our identification method still yields moderate incompleteness highlights the importance of conducting follow-up interferometric observations to provide completely reliable sample of SMGs \citep{{Hodge:2013lr}}.} We find that the identification rate is lower for fainter SCUBA-2 sources, and argue that for follow-up observations sensitive to SMGs with $S_{850}\sim1$\,mJy across the whole ALMA beam, the multiple fraction is likely to be $\gtrsim40$\% for sources with $S_{850} \gtrsim 4$\,mJy.

\item The redshift distribution based on the {photometric redshifts of the} candidate SMG counterparts of the $\geq4\,\sigma$ {\it Class} = 1 SCUBA-2 sources is well fit by a log-normal distribution, with a median redshift of $z=2.3\pm0.1$ (as found by \citet{Simpson:2014aa}). Based on the selection curves and the FIR colors, we argue that submillimeter sources without any identification are likely to be located at $z \gtrsim 3$. After accounting for these unidentified sources, we estimate that the median redshift for SMGs with $S_{850} > 1$\,mJy to be $z=2.6\pm0.1$. Our results are in good agreement with model predictions and previous observational measurements.  

\item Using the Landy \& Szalay estimator we find a strong angular clustering signal, although still with large uncertainties, for candidate SMGs associated with $\geq4\,\sigma$ {\it Class} = 1 SCUBA-2 sources at $1<z<3$. The clustering signal roughly corresponds to a correlation length of $r_0 = 21^{+6}_{-7} h^{-1}$Mpc, or galaxy bias of $b = 9.1^{+2.1}_{-2.8}$, and a DM halo mass of $M_{halo} = 8\pm5 \times10^{13} h^{-1} M_\odot$. Our results highlight the fact that larger samples of SMGs with better determinations on redshifts, or cross-correlation with other populations with larger sample size, are needed to provide better constraints on the SMG clustering properties, and so test the relationship of this population to local galaxies.

\end{enumerate}
\section*{Acknowledgments}
{We thank the referee for a helpful report that has improved the manuscript.}
We would like to thank Peder Norberg for helpful discussions on the clustering measurement. C.-C.C. is especially grateful to Tzu-Ying Lee (李姿瑩), who ensures the arrival of Han-Ching Neo Chen (陳翰青). C.-C.C., I.R.S. acknowledge support from the ERC Advanced Investigator programme DUSTYGAL 321334. I.R.S. also acknowledges support from a Royal Society/Wolfson Merit Award and STFC through grant number ST/L00075X/1. A.M.S. acknowledges financial support from an STFC Advanced Fellowship (ST/H005234/1) \& Leverhulme foundation. 
The James Clerk Maxwell Telescope has historically been operated by the Joint Astronomy Centre on behalf of the Science and Technology Facilities Council of the United Kingdom, the National Research Council of Canada and the Netherlands Organisation for Scientific Research. Additional funds for the construction of SCUBA-2 were provided by the Canada Foundation for Innovation. This paper makes use of the following ALMA data: ADS/JAO.ALMA\#2012.1.00090.S. ALMA is a partnership of ESO (representing its member states), NSF (USA) and NINS (Japan), together with NRC (Canada), NSC and ASIAA (Taiwan), and KASI (Republic of Korea), in cooperation with the Republic of Chile. The Joint ALMA Observatory is operated by ESO, AUI/NRAO and NAOJ.
This research made use of Astropy, a community-developed core Python package for Astronomy \citep{Astropy-Collaboration:2013aa}. This research has made use of NASA's Astrophysics Data System. The authors wish to recognize and acknowledge the very significant cultural role and reverence that the summit of Mauna Kea has always had within the indigenous Hawaiian community. We are most fortunate to have the opportunity to conduct observations from this mountain.

\end{CJK}

\bibliography{bib}

\appendix
The selected candidate counterparts for the {\sc main} (SNR $\geq\,4\,\sigma$) and {\sc supplementary} ($3.5\,\sigma \leq$ SNR $\leq 4.0\,\sigma$) SCUBA-2 sources are shown in \autoref{tab:ids} and \autoref{tab:idssup}, respectively. Column 1, 2, and 3 are short IDs, R.A., and Decl. in degrees of the SCUBA-2 sources, Column 4, 5, and 6 give for each SCUBA-2 source its {\it Class} (see \autoref{subsec:id} for details about the classification), the total number of the counterpart candidates, and the assigned IDs for each counterpart candidate, Column 7, 8 (9, 10) show the R.A. and Decl. in degrees of each radio ($K$-band) counterpart, if any, Column 11 gives the photometric redshifts with errors, Column 12 shows the $p$-value for counterparts that are selected by radio, and Column 13 shows the mean SMG fraction of candidates selected purely based on the OIRTC technique ($\langle f_{\rm OIRTC}\rangle$; see \autoref{subsec:oir} for the definition of this quantity). Note that the candidate counterparts that are confirmed by the ALMA observations \citep{Simpson:2015ab} are denoted with $^\ast$ in Column 6. In addition \citet{Simpson:2015ab} detect additional SMGs in the following fields (coordinates and properties are given in Table 1 of \citet{Simpson:2015ab}:  UDS.0003, UDS.0004, UDS.0005, UDS.0007, UDS.0017, UDS.0020, UDS.0023, UDS.0024, UDS.0033, UDS.0047, UDS.0051, UDS.0078.
\LongTables
\setlength{\tabcolsep}{0.033in}
\begin{deluxetable*}{lcccccrrccccc}
\tablewidth{10in}
\tablecaption{Candidate SMG counterparts of the {\sc main} SCUBA-2 sources\label{tab:ids}}
\tablehead{
\colhead{S2CLS ID} & \colhead{SMG R.A.} & \colhead{SMG Decl.} & \colhead{{\it Class}} & \colhead{\#ID} & \colhead{ID\_ID} & \colhead{Radio R.A.} & \colhead{Radio Decl.} & \colhead{$K$-band R.A.} & \colhead{$K$-band Decl.} & \colhead{$z_{\rm photo}$} & \colhead{$p$} & \colhead{$\langle f_{\rm OIRTC}\rangle$} \\
 \colhead{} & \colhead{[Degree]} & \colhead{[Degree]} & \colhead{} & \colhead{} & \colhead{} & \colhead{[Degree]} & \colhead{[Degree]} & \colhead{[Degree]} & \colhead{[Degree]} & \colhead{} & \colhead{} & \colhead{} \\
 \colhead{(1)} & \colhead{(2)} & \colhead{(3)} & \colhead{(4)} & \colhead{(5)} & \colhead{(6)} & \colhead{(7)} & \colhead{(8)} & \colhead{(9)} & \colhead{(10)} & \colhead{(11)} & \colhead{(12)} & \colhead{(13)} }
\startdata
UDS.0001 & 34.62821 & $-$5.52522 & 1 & 1 & ID1 & 34.62779 & $-$5.52542 & 34.62772 & $-$5.52550 & 1.46$^{+0.08}_{-0.09}$ & 0.001 & \ldots\\
UDS.0002 & 34.60139 & $-$5.38246 & 1 & 2 & ID1$^\ast$ & 34.60079 & $-$5.38225 & 34.60096 & $-$5.38252 & 2.32$^{+0.54}_{-0.36}$ & 0.005 & \ldots\\
 &  &  &  &  & ID2 & \ldots & \ldots & 34.60195 & $-$5.38018 & 3.59$^{+1.25}_{-0.55}$ & \ldots & 0.41\\
UDS.0003 & 34.83828 & $-$4.94792 & 1 & 2 & ID1$^\ast$ & 34.83821 & $-$4.94767 & 34.83803 & $-$4.94742 & 1.31$^{+0.48}_{-0.10}$ & 0.004 & \ldots\\
 &  &  &  &  & ID2$^\ast$ & \ldots & \ldots & 34.83917 & $-$4.94688 & 2.64$^{+0.23}_{-0.36}$ & \ldots & 0.31\\
UDS.0004 & 34.20033 & $-$5.02520 & 1 & 1 & ID1$^\ast$ & 34.19967 & $-$5.02492 & 34.19973 & $-$5.02490 & 3.10$^{+0.22}_{-0.28}$ & 0.019 & \ldots\\
UDS.0005 & 34.35754 & $-$5.42691 & 1 & 1 & ID1$^\ast$ & 34.35725 & $-$5.42819 & 34.35732 & $-$5.42826 & 0.44$^{+0.02}_{-0.05}$ & 0.025 & \ldots\\
UDS.0006 & 34.52381 & $-$5.18080 & 1 & 1 & ID1$^\ast$ & 34.52354 & $-$5.18039 & 34.52376 & $-$5.18043 & 3.59$^{+0.40}_{-0.28}$ & 0.009 & \ldots\\
UDS.0007 & 34.37708 & $-$5.32302 & 1 & 1 & ID1$^\ast$ & 34.37688 & $-$5.32289 & \ldots & \ldots & \ldots & 0.005 & \ldots\\
UDS.0008 & 34.51268 & $-$5.47858 & 1 & 2 & ID1$^\ast$ & 34.51250 & $-$5.47833 & 34.51251 & $-$5.47821 & 3.46$^{+0.47}_{-0.20}$ & 0.004 & \ldots\\
\enddata
\tablecomments{Table 3 will be published in its entirety in the electronic 
edition of the {\it Astrophysical Journal}.  A portion is shown here 
for guidance regarding its form and content.}
\end{deluxetable*}

\LongTables
\setlength{\tabcolsep}{0.033in}
\begin{deluxetable*}{lcccccrrccccc}
\tablewidth{10in}
\tablecaption{Candidate SMG counterpars of the {\sc supplementary} SCUBA-2 sources\label{tab:idssup}}
\tablehead{
\colhead{S2CLS ID} & \colhead{SMG R.A.} & \colhead{SMG Decl.} & \colhead{{\it Class}} & \colhead{\#ID} & \colhead{ID\_ID} & \colhead{Radio R.A.} & \colhead{Radio Decl.} & \colhead{$K$-band R.A.} & \colhead{$K$-band Decl.} & \colhead{$z_{\rm photo}$} & \colhead{$p$} & \colhead{$\langle f_{\rm OIRTC}\rangle$} \\
 \colhead{} & \colhead{[Degree]} & \colhead{[Degree]} & \colhead{} & \colhead{} & \colhead{} & \colhead{[Degree]} & \colhead{[Degree]} & \colhead{[Degree]} & \colhead{[Degree]} & \colhead{} & \colhead{} & \colhead{} \\
 \colhead{(1)} & \colhead{(2)} & \colhead{(3)} & \colhead{(4)} & \colhead{(5)} & \colhead{(6)} & \colhead{(7)} & \colhead{(8)} & \colhead{(9)} & \colhead{(10)} & \colhead{(11)} & \colhead{(12)} & \colhead{(13)} }
\startdata
UDS.0714 & 34.65998 & $-$5.36077 & 1 & 0 & \ldots & \ldots & \ldots & \ldots & \ldots & \ldots & \ldots & \ldots\\
UDS.0715 & 33.98784 & $-$5.04675 & 2 & 0 & \ldots & \ldots & \ldots & \ldots & \ldots & \ldots & \ldots & \ldots\\
UDS.0716 & 34.79200 & $-$4.95184 & 1 & 0 & \ldots & \ldots & \ldots & \ldots & \ldots & \ldots & \ldots & \ldots\\
UDS.0717 & 34.68288 & $-$5.43243 & 1 & 1 & ID1 & \ldots & \ldots & 34.68155 & $-$5.43050 & 1.58$^{+0.38}_{-0.07}$ & \ldots & 0.12\\
UDS.0718 & 34.40001 & $-$4.71081 & 1 & 1 & ID1 & 34.39950 & $-$4.70931 & 34.39982 & $-$4.70935 & \ldots & 0.065 & \ldots\\
UDS.0719 & 34.08140 & $-$5.26570 & 1 & 0 & \ldots & \ldots & \ldots & \ldots & \ldots & \ldots & \ldots & \ldots\\
UDS.0720 & 34.64493 & $-$5.42466 & 1 & 5 & ID1 & 34.64512 & $-$5.42467 & 34.64504 & $-$5.42456 & \ldots & 0.004 & \ldots\\
 &  &  &  &  & ID2 & \ldots & \ldots & 34.64428 & $-$5.42475 & 1.72$^{+0.51}_{-0.03}$ & \ldots & 0.18\\
 &  &  &  &  & ID3 & 34.64492 & $-$5.42592 & 34.64489 & $-$5.42591 & 1.65$^{+0.22}_{-0.10}$ & 0.018 & \ldots\\
 &  &  &  &  & ID4 & 34.64342 & $-$5.42486 & 34.64332 & $-$5.42485 & 1.07$^{+0.06}_{-0.04}$ & 0.023 & \ldots\\
\enddata
\tablecomments{Table 4 will be published in its entirety in the electronic 
edition of the {\it Astrophysical Journal}.  A portion is shown here 
for guidance regarding its form and content.}
\end{deluxetable*}

\makeatletter 
\renewcommand{\thefigure}{A\@arabic\c@figure}
\makeatother
%
%
\setcounter{figure}{0}
\begin{figure*}
      \includegraphics[scale=2]{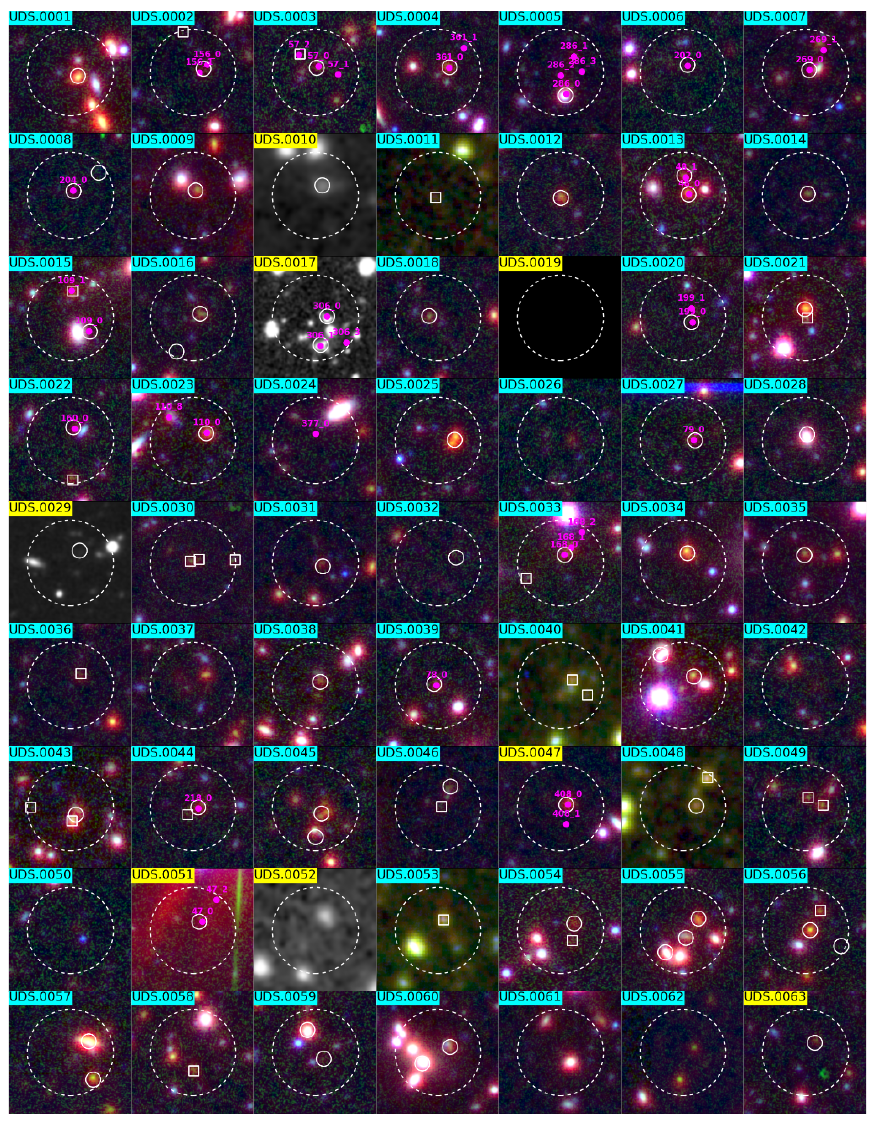}
       \caption{False-color Ch1-$K$-$z$ (r-g-b) thumbnails (Ch2-Ch1-$K$ if $z$-band is not available) for the {\sc main} SCUBA-2 sources. If less than three photometric bands are available, we show in grey scale the image of the shortest among the three waveband available. The size of each box is 25$'' \times 25''$ and the large dashed circles show the counterpart searching area with a radius of $8\farcs7$. The solid squares and circles mark the counterpart candidates identified through the OIRTC technique and radio imaging, respectively. We only show circles if sources are identified by both radio and OIRTC. The magenta points are ALMA-detected SMGs with the ID numbers adopted from \citet{Simpson:2015ab}. The background color for each ID number shows the {\it Class} of each source, with cyan, yellow, and brown marking {\it Class} =1, 2, and 3, respectively (matched to the color scheme adopted in \autoref{multicover}). The complete version of this figure will be published in the electronic edition of the {\it Astrophysical Journal}.}
    \label{postage_main}
\end{figure*}

\setcounter{figure}{1}
\begin{figure*}
      \includegraphics[scale=2]{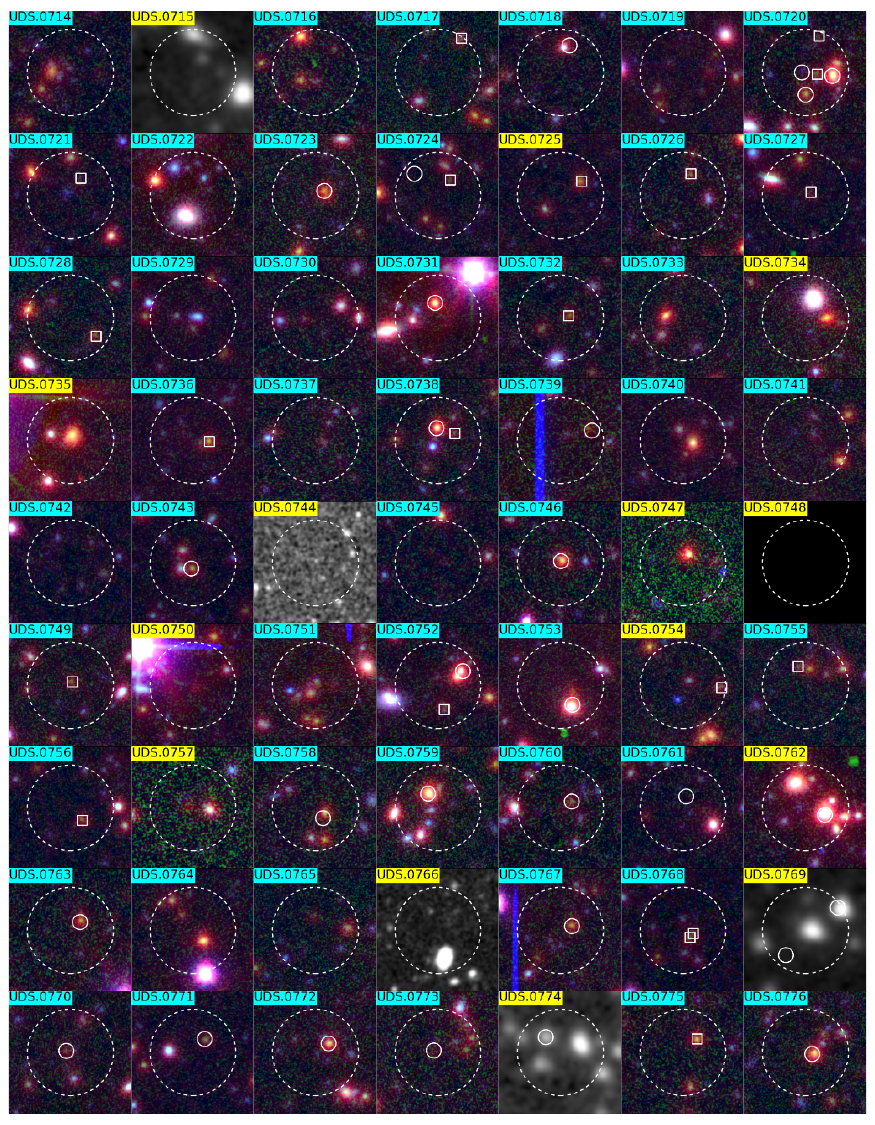}
       \caption{Same as \autoref{postage_main}, but on the {\sc supplementary} sources. The complete version of this figure will be published in the electronic edition of the {\it Astrophysical Journal}.}
    \label{postage_suppl}
\end{figure*}

\end{document}